\documentclass[epj]{svjour} 

\usepackage{graphicx}
\usepackage{epsfig}
\usepackage{amssymb}
\usepackage{pslatex}

\newcommand{\sqrtsnn}{\sqrt{s_{_{NN}}}}
\def\mean#1{\ensuremath{\left<#1\right>}}

\begin{document}

\title{Probing the QCD equation of state with thermal photons in nucleus-nucleus collisions at RHIC}
\author{David d'Enterria\inst{\star}, and Dmitri Peressounko\inst{\dagger}
}                     
\institute{
$^\star$Nevis Laboratories, Columbia University, Irvington, NY 10533, and New York, NY 10027, USA\\
$^\dagger$ RRC ``Kurchatov Institute'', Kurchatov Sq. 1, Moscow, 123182
}

\date{Received: date / Revised version: date}

\abstract{
Thermal photon production at mid-rapidity in Au+Au reactions at
$\sqrt{s_{\mbox{\tiny{\it{NN}}}}}$ = 200 GeV is studied in the
framework of a hydrodynamical model that describes 
efficiently the bulk identified hadron spectra at RHIC.
The combined thermal plus NLO pQCD photon spectrum is in good
agreement with the yields measured by the PHENIX experiment for 
all Au+Au centralities. Within our model, we demonstrate that
the correlation of the thermal photon slopes with the
charged hadron multiplicity in each centrality provides direct
empirical information on the underlying degrees of freedom and on the form 
of the equation of state, $s(T)/T^3$, of the strongly interacting matter
produced in the course of the reaction.
} 

\titlerunning{Probing the QCD EoS with thermal $\gamma$ in A+A collisions at RHIC}
\maketitle


\section{Introduction}

Numerical calculations of lattice QCD predict a transition from
ordinary hadronic matter to a deconfined state of quarks and gluons when
the temperature of the 
system is of the order of $T_{crit}\approx$ 0.17 GeV~\cite{latt}.
The existence of such a phase transition manifests itself clearly
in the QCD equation-of-state (EoS) on the lattice by a sharp jump of
the (Stefan-Boltzmann) scaled energy density, $\varepsilon(T)/T^4$,
at the critical temperature, reminiscent of a first-order phase
change\footnote{The order of the phase transition itself is not
exactly known: the pure SU(3) gauge theory is first-order whereas
introduction of 2+1 flavours makes it of a fast cross-over
type~\cite{latt}.}. The search for evidences of this deconfined
plasma of quarks and gluons (QGP) is the main driving
force behind the study of relativistic nuclear collisions 
at different experimental facilities in the last 20 years. Whereas several
experimental results have been found consistent with the formation
of the QGP both at CERN-SPS~\cite{sps_qgp} and BNL-RHIC~\cite{rhic_qgp}
energies, it is fair to acknowledge that there is no incontrovertible proof
yet of bulk deconfinement in the present nucleus-nucleus data. In this paper,
we present a detailed study of the only experimental signature, thermal photons,
that can likely provide direct information on the {\it thermodynamical}
properties (and, thus, on the equation-of-state) of the underlying QCD
matter produced in high-energy heavy-ion collisions.
Electromagnetic radiation (real and virtual photons) emitted in the course
of a heavy-ion reaction, has long~\cite{feinberg,shuryak_photons} been considered
a privileged probe of the space-time evolution of the colliding 
system\footnote{Excellent reviews on photon production in relativistic nuclear collisions
have been published recently~\cite{peitz_thoma_physrep,yellow_rep,gale_rep}.},
inasmuch as photons are not distorted by final-state interactions due
to their weak interaction with the surrounding medium. Direct photons,
defined as real photons not originating from the decay of final hadrons, are
emitted at various stages of the reaction with several contributing processes.
Two generic mechanisms are usually considered:
(i) {\it prompt} (pre-equilibrium or pQCD) photon emission from perturbative
parton-parton scatterings in the first tenths of fm/$c$ of the collision process,
(ii) subsequent $\gamma$ emission from the {\it thermalized} partonic (QGP) and hadronic
(hadron resonance gas, HRG) phases of the reaction.\\

\begin{sloppypar}
Experimentally, direct $\gamma$ have been indeed measured in Pb+Pb
collisions at CERN-SPS ($\sqrt{s_{\mbox{\tiny{\it{NN}}}}}$ = 17.3
GeV)~\cite{wa98_photons}. However, the relative contributions to
the total spectrum of the pQCD, QGP and HRG components have not
been determined conclusively. Different hydrodynamics
calculations~\cite{srivastava_sps_rhic,alam_sps_rhic,peressou,steffen_sps_rhic_lhc,finnish_hydro}
require ``non-conventional'' conditions: high initial temperatures
($T_{0}^{max}>$ $T_{crit}$), strong partonic and/or hadronic
transverse velocity flows, or in-medium modifications of hadron
masses, in order to reproduce the observed photon spectrum.
However, no final conclusion can be drawn from these results due
mainly to the uncertainties in the exact amount of radiation
coming from primary parton-parton collisions. In a situation akin
to that affecting the interpretation of high $p_T$ hadron data at
SPS~\cite{dde_sps}, the absence of a concurrent baseline
experimental measurement of prompt photon production in p+p
collisions at the same $\sqrt{s}$ and $p_T$ range as the
nucleus-nucleus data, makes it difficult
to have any reliable empirical estimate of the actual
thermal $\gamma$ excess in the Pb+Pb spectrum. In the theoretical
side, the situation at SPS is not fully under control either: (i)
next-to-leading-order (NLO) perturbative calculations are known to
underpredict the experimental reference nucleon-nucleon $\gamma$
differential cross-sections below $\sqrt{s}\approx$
30~GeV~\cite{aurenche} (a substantial amount of parton
intrinsic transverse momentum $k_T$~\cite{wong}, approximating the
effects of parton Fermi motion and soft gluon radiation, is
required~\cite{apanasevich}), (ii) the implementation of the extra nuclear $k_T$
broadening  observed in the nuclear data (``Cronin
enhancement''~\cite{cronin} resulting from multiple soft and
semi-hard interactions of the colliding partons on their way
in/out the traversed nucleus) is
model-dependent~\cite{dumitru,ina,levai} and introduces an additional
uncertainty to the computation of the yields, and (iii)
hydrodynamical calculations usually assume initial conditions
(longitudinal boost invariance, short thermalization times, zero
baryochemical potential) too idealistic for SPS energies. The
situation at RHIC (and LHC) collider energies is undoubtedly far
more advantageous. Firstly, the photon spectra for different
centralities in Au+Au~\cite{ppg042} and in (baseline) p+p~\cite{ppg049}
collisions at $\sqrt{s}$ = 200 GeV are already experimentally available.
Secondly, the p+p baseline reference is well under control
theoretically (NLO calculations do not require extra
non-perturbative effects to reproduce the hard spectra at
RHIC~\cite{ppg049,ppg024}). Thirdly, the amount of nuclear
Cronin enhancement experimentally observed is very modest
(high $p_T$ $\pi^0$ are barely enhanced in d+Au collisions at
$\sqrt{s_{NN}}$ = 200 GeV~\cite{dAu_phnx}), and one expects even
less enhancement for $\gamma$ which, once produced, do not gain
any extra $k_T$ in their way out through the nucleus.
Last but not least, the produced system at midrapidity in heavy-ion
reactions at RHIC top energies is much closer to the zero net baryon
density and longitudinally boost-invariant conditions customarily
presupposed in the determination of the parametrized photon rates
and in the hydrodynamical implementations of the reaction evolution.
In addition, the thermalization times usually assumed in the hydrodynamical
models ($\tau_{\mbox{\tiny{\it{therm}}}}\lesssim$ 1 fm/$c$) are, for the
first time at RHIC, above the lower limit imposed by the transit time of
the two colliding nuclei ($\tau_0 = 2R/\gamma\approx$ 0.15 fm/$c$ for Au+Au at 200 GeV).
As a matter of fact, it is for the first time at RHIC that hydrodynamics
predictions agree {\it quantitatively} with most of the differential observables
of bulk (``soft'') hadronic production below $p_T\approx$ 1.5 GeV/$c$ in
Au+Au reactions~\cite{kolb_heinz_rep,teaney_hydro,hirano}.\\
\end{sloppypar}

\begin{sloppypar}
In this context, the purpose of this paper is three-fold.
First of all, we present a relativistic Bjorken hydrodynamics model that
reproduces well the identified hadron spectra measured  at all centralities
in Au+Au collisions at $\sqrt{s_{NN}}$ = 200 GeV (and, thus, the 
centrality dependence of the total charged hadron multiplicity).
Secondly, using such a model complemented with the most up-to-date
parametrizations of the QGP and HRG photon emission rates, we determine the
expected thermal photon yields in Au+Au reactions and compare them to the
prompt photon yields computed in NLO perturbative QCD. The combined inclusive
(thermal+pQCD) photon spectrum is successfully confronted to recent results
from the PHENIX collaboration as well as to other available predictions.
Thirdly, 
after discussing in which $p_T$ range the thermal photon signal
can be potentially identified experimentally, we address the issue
of how to have access to the thermodynamical properties (temperature,
entropy density) of the radiating matter. We propose the correlation of two
experimentally measurable quantities: the thermal photon slope and
the multiplicity of charged hadrons produced in the reaction,
as a direct method to determine the underlying degrees of freedom 
and the equation of state, $s(T)/T^3$, of the dense and hot QCD
medium produced in Au+Au collisions at RHIC energies.
\end{sloppypar}

\section{Hydrodynamical model}

\subsection{Implementation} 

\begin{sloppypar}
Hydrodynamical approaches of particle production in heavy-ion 
collisions assume {\it local} conservation of energy and momentum 
in the hot and dense strongly interacting matter produced in the course 
of the reaction and describe its evolution using the equations of motion of 
perfect (non-viscous) relativistic hydrodynamics. These equations are nothing 
but the conservation of:
\begin{description}
\item (i) the energy-momentum tensor: $\partial_{\mu}
T^{\mu\nu} = 0$ with $T^{\mu\nu} = (\varepsilon +
p)u^{\mu}u^{\nu}-p\,g^{\mu\nu}$ [where $\varepsilon$, $p$, and
$u^{\nu}=(\gamma,\gamma$v) are resp. the energy density,
pressure, and collective flow 4-velocity fields, and
$g^{\mu\nu}$=diag(1,-1,-1,-1) the metric tensor], and 
\item (ii) the conserved currents in strong interactions: $\partial_\mu
J^{\mu}_{i} = 0$, with $J^{\mu}_{i}=n_{i}u^{\mu}$ [where $n_i$ is
the number density of the net baryon, electric charge, net strangeness,
etc. currents].
\end{description}
\end{sloppypar}

These equations complemented with three input ingredients:
(i) the initial conditions ($\varepsilon_0$ 
at time $\tau_0$),
(ii) the equation-of-state of the system, $p(\varepsilon,n_{i})$,
relating the local thermodynamical quantities, 
and (ii) the freeze-out conditions, describing the transition from
the hydrodynamics regime to the free streaming final particles,
are able to reproduce most of the bulk hadronic observables
measured in heavy-ion reactions at RHIC~\cite{kolb_heinz_rep,teaney_hydro,hirano}.\\

\begin{sloppypar}
The particular hydrodynamics implementation used in this work is
discussed in detail in~\cite{peressou}. 
We assume cylindrical symmetry in the transverse direction ($r$) 
and longitudinal ($z$) boost-invariant (Bjorken) expansion~\cite{bjorken}
which reduces the equations of motion to a one-dimensional problem
but results in a loss of the dependence of
the observables on longitudinal degrees of freedom. Our results,
thus, are only relevant for particle production within a finite range
around midrapidity\footnote{The experimental $\pi^\pm$ and $K^\pm$
$dN/dy$ distributions at RHIC are Gaussians~\cite{brahms_hadrons},
as expected from perturbative QCD initial conditions~\cite{eskola_hydro}.
Thus, although there is no Bjorken rapidity plateau, the widths of the distributions
are quite broad and within $|y|\lesssim$ 2, deviations from boost
invariance are not very large~\cite{eskola_hydro}.}.
The equation-of-state used here describes a first order phase
transition from a QGP to a HRG at $T_{crit}$ = 165 MeV with latent
heat\footnote{Although the lattice results seem to indicate that
the transition is of a fast cross-over type, the predicted change
of $\Delta\varepsilon \approx$ 0.8 GeV/fm$^3$ in a narrow temperature
interval of $\Delta T\approx$ 20 MeV~\cite{latt} can be interpreted as the latent
heat of the transition.} $\Delta\varepsilon\approx$ 1.4 GeV/fm$^3$,
very similar to that used in other works~\cite{kolb_heinz_rep}.
The QGP is modeled as an ideal gas of massless quarks ($N_f$
= 2.5 flavours) and gluons with total degeneracy
$g_{\mbox{\tiny{\it{QGP}}}} = (g_{\mbox{\tiny{\it{gluons}}}}+7/8\,
g_{\mbox{\tiny{\it{quarks}}}})$ = 42.25. The corresponding EoS,
$p=1/3\varepsilon-4/3B$ ($B$ being the bag constant), has sound
velocity $c_s^2=\partial p/\partial \varepsilon = 1/3$.
The hadronic phase is modeled as a non-interacting gas of
$\sim$400 known hadrons and hadronic resonances with masses below
2.5 GeV/$c^2$. The inclusion of heavy hadrons leads to an equation
of state significantly different than that of an ideal gas of
massless pions: the velocity of sound in the HRG phase is
$c_s^2\approx$ 0.15, resulting in a relatively soft hadronic EoS
as suggested by lattice calculations~\cite{mohanty_cs}; and the
effective number of degrees of freedom at $T_{c}$ is
$g_{\mbox{\tiny{\it{HRG}}}}\approx$ 12 (as given by $g_{\ensuremath{\it eff}} =
45\,s/(2\pi^2\,T^3)$, see later). Both phases are
connected via the standard Gibbs' condition of phase equilibrium, 
$p_{\mbox{\tiny{\it{QGP}}}}(T_{c}) =
p_{\mbox{\tiny{\it{HRG}}}}(T_{c})$, during the mixed phase.
The external bag pressure, calculated to fulfill this condition at
$T_c$, is $B\approx$ 0.38 GeV/fm$^3$. The system of equations is solved
with the MacCormack two-step (predictor-corrector) numerical scheme~\cite{maccormack}
with time and radius steps: $\delta t$ = 0.02 fm/$c$ and $\delta r$ = 0.1 fm 
respectively.\\
\end{sloppypar}

Statistical model analyses of particle production in nucleus-nucleus
reactions~\cite{pbm_thermal} provide a very good description of
the measured particle ratios at RHIC assuming that all hadrons are
emitted from a thermalized system reaching chemical equilibrium at
a temperature $T_{chem}$ with baryonic, strange and isospin chemical
potentials $\mu_{i}$. In agreement with those observations, our specific
hydrodynamical evolution reaches chemical freeze-out at $T_{chem}=150$ MeV
with $\mu_{B}=25$ MeV (as given by the latest statistical fits to hadron 
ratios~\cite{andronic05}), and has $\mu_{S}=\mu_{I}=0$. For temperatures above $T_{chem}$ 
we conserve baryonic, strange and charge currents, but not particle numbers, 
while for temperatures below $T_{chem}$ we explicitly conserve particle
numbers by introducing individual (temperature-dependent)
chemical potentials for each hadron.
The final differential hadron $dN/dp_T$ spectra are produced via a standard
Cooper-Frye ansatz~\cite{cooper_frye} at the kinetic freeze-out
temperature ($T_{\ensuremath{\it fo}}=120$ MeV) when the hydrodynamical
equations lose their validity, i.e. when the microscopic length (the
hadrons mean free path) is no longer small compared to the size of the system.
Unstable resonances are then allowed to decay with their appropriate branching
ratios~\cite{PDG}. Table I summarizes the most important
parameters describing our hydrodynamic evolution.
The only free parameters are the initial energy density $\varepsilon_0$
in the center of the reaction zone for head-on (impact parameter $b$ = 0 fm) 
Au+Au collisions at the starting time $\tau_0$, and the temperature at
freeze-out time, $T_{\ensuremath{\it fo}}$.

\subsection{Initialization}

\begin{figure}[htbp]
\begin{center}
 \psfig{figure=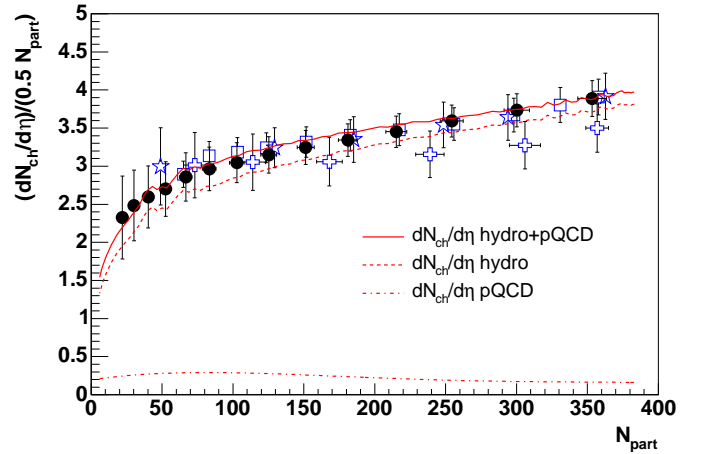,width=9.cm}
\end{center}
\caption{Charged hadron multiplicity at midrapidity (normalized by the 
number of participant nucleon pairs) as a function of centrality (given by 
the number of participants, $N_{part}$) measured in Au+Au 
at $\sqrt{s_{\mbox{\tiny{\it{NN}}}}}$ = 200 GeV by PHENIX~\protect\cite{ppg019} 
(circles), STAR~\protect\cite{star_Nch} (stars), PHOBOS~\protect\cite{phobos_Nch} 
(squares) and BRAHMS~\protect\cite{brahms_Nch} (crosses), compared to our 
hydrodynamics calculations (dashed line), our scaled pQCD ($p_T>$ 1 GeV/$c$) 
p+p yields~\protect\cite{vogel_hadrons} (dashed-dotted line), and to the sum 
hydro+pQCD (solid line).}
\label{fig:dNch}
\end{figure}

\begin{sloppypar}
We distribute the initial energy density within the reaction volume 
according to the geometrical Glauber\footnote{The density of participant 
and colliding nucleons are obtained from the nuclear overlap function 
$T_{AA}(b)$ computed with a Glauber Monte Carlo code which parametrizes
the Au nuclei with Woods-Saxon functions with radius $R$ = 6.38~fm and 
diffusivity $a$ = 0.54~fm~\cite{hahn}.} 
prescription proposed by Kolb {\it et al.}~\cite{kolb_heinz_finnishgroup}. 
Such an ansatz ascribes 75\% of the initial entropy production in a given 
centrality bin, $s_0(b)$, to soft processes (scaling with the transverse 
density of participant nucleons $N_{part}(b)$) and the remaining 25\% to 
hard processes (scaling with the density of point-like collisions, 
$N_{coll}(b)$, proportional to the nuclear overlap function $T_{AA}(b)$):
\begin{equation}
s(b) = C\cdot(0.25\cdot N_{part}(b) + 0.75 \cdot N_{coll}(b)),
\end{equation}
where $C$ is a normalization coefficient chosen so that we produce the correct 
particle multiplicity at $b$ = 0 fm. 
For each impact parameter, we construct an azimuthally symmetric hydrodynamical 
source from the (azimuthally deformed) initial Glauber entropy distribution, 
by defining a coordinate origin in the middle point between the centers of 
the two colliding nuclei and averaging the entropy density over all azimuthal 
directions. We then transform $\varepsilon_0(b)\propto s_0(b)^{4/3}$.
This method provides a very good description of the measured
centrality dependence of the final charged hadron rapidity 
densities $dN_{ch}/d\eta$ measured at RHIC as can be seen in Figure~\ref{fig:dNch}. 
Note that in our implementation of this prescription, we explicitly added 
the contribution of the particle multiplicity coming from hard processes 
(i.e. from hadrons having $p_T>$ 1 GeV/$c$) obtained from the scaled pQCD 
calculations (see later). Such a ``perturbative'' component accounts for a 
roughly constant $\sim$7\% factor of the total hadron multiplicity for all centralities. 
The good reproduction of the measured charged hadron integrated yields is an 
important result for our later use of $dN_{ch}/d\eta|_{\eta=0}$ as an empirical 
measure of the initial entropy density in different Au+Au centrality classes 
(see Section~\ref{sec:eos}).\\
\end{sloppypar}

\begin{table*}[htb]
\caption{Summary of the thermodynamical parameters characterizing
our hydrodynamical model evolution for central ($b = 0$ fm) Au+Au
collisions at $\sqrt{s_{\mbox{\tiny{\it{NN}}}}}$ = 200 GeV. Input
parameters are the (maximum) initial energy density $\varepsilon_0$
(with corresponding ideal-gas entropy densities $s_0$ and 
temperature $T_0$)
at time $\tau_0$, the baryochemical potential $\mu_{B}$, and the chemical and 
kinetic freeze-out temperatures $T_{chem}$ and $T_{\ensuremath{\it fo}}$ (or 
energy density $\varepsilon_{\ensuremath{\it fo}}$). The energy densities at the end 
of the pure QGP ($\varepsilon_{\mbox{\tiny{\it{QGP}}}}^{\mbox{\tiny{\it{min}}}}$),
and at the beginning of the pure hadron gas phase
($\varepsilon_{\mbox{\tiny{\it{HRG}}}}^{\mbox{\tiny{\it{max}}}}$) are
also given for indication, as well as the average (over total volume) values of the
initial energy density $\langle\varepsilon_0\rangle$, entropy density $\langle s_0\rangle$, 
and temperature $\langle T_0\rangle$.}
\begin{center}
\begin{tabular}{c|c|c|c|c|c|c|c|c|c}
\hline\hline $\tau_0$ & $\varepsilon_0$ ($\langle\varepsilon_0\rangle)$
& $s_0$ ($\langle s_0\rangle$)
& $T_0$ ($\langle T_0\rangle$) &
$\varepsilon_{\mbox{\tiny{\it{QGP}}}}^{\mbox{\tiny{\it{min}}}}$ &
$\varepsilon_{\mbox{\tiny{\it{HRG}}}}^{\mbox{\tiny{\it{max}}}}$ &
$\mu_{B}$ &
$T_{chem}$ & $T_{\ensuremath{\it fo}}$ & $\varepsilon_{\ensuremath{\it fo}}=\varepsilon_{\mbox{\tiny{\it{HRG}}}}^{\mbox{\tiny{\it{min}}}}$ \\
(fm/$c$) & (GeV/fm$^3$) & (fm$^{-3}$) & 
(MeV) & (GeV/fm$^3$) & (GeV/fm$^3$) & (MeV) & (MeV) & (MeV) & (GeV/fm$^3$)
\\\hline
0.15 & 220 (72) &  498 (190) & 590 (378) & 1.7 & 0.35 & 25. & 150 & 120 & 0.10 \\
\hline\hline
\end{tabular}
\label{tab:hydro_parameters}
\end{center}
\end{table*}

For the initial conditions (Table~\ref{tab:hydro_parameters}),
we choose $\varepsilon_0$ = 220 GeV/fm$^3$ (maximum energy density at
$b$ = 0 fm, corresponding to an {\it average} energy density over
the total volume for head-on collisions of
$\langle\varepsilon_0\rangle$ = 72 GeV/fm$^3$) at a
time $\tau_0 = 2R/\gamma\approx$ 0.15 fm/$c$ equal to the transit time
of the two Au nuclei at $\sqrt{s_{\mbox{\tiny{\it{NN}}}}}$ = 200 GeV.
The choice of this relatively short value of $\tau_0$, -- otherwise typically
considered in other hydrodynamical studies of thermal photon
production at RHIC~\cite{srivastava_sps_rhic,finnish_hydro,frankfurt_rhic_lhc}
--, rather than the ``standard'' thermalization time of $\tau_{therm}$
= 0.6 fm/$c$~\cite{kolb_heinz_rep,teaney_hydro,hirano}, is driven by our will to
consistently take into account within our space-time evolution the
emission of photons from secondary ``cascading'' parton-parton
collisions~\cite{bass,bass2} taking place in the {\it thermalizing} phase
between prompt pQCD emission (at $\tau\sim 1/p_T\lesssim$0.15 fm/$c$) and full 
equilibration (see Sect.~\ref{sec:extra_gamma}). Though it may be questionable
to identify such photons from second-chance parton-parton collisions as 
genuine {\it thermal} $\gamma$, it is clear that their spectrum reflects the 
momentum distribution of the partons during this thermalizing phase\footnote{Note
also that it is precisely those secondary partonic interactions that are actually 
driving the system towards (local) thermal equilibrium.}. Additionally, recent 
theoretical works~\cite{berges04,arnold04} do seem to support the application 
of hydrodynamics equations in such ``pre-thermalization'' conditions. 
Our consequent space-time evolution leads to a value of the energy density
of $\varepsilon\approx$ 30 GeV/fm$^3$ at $\tau_{therm}$ = 0.6 fm/$c$,
in perfect agreement with other 2D+1 hydrodynamic calculations which do not
invoke azimuthal symmetry~\cite{kolb_heinz_rep,teaney_hydro}
as well as more numerically involved 3D+1 approaches~\cite{hirano}.
Thus, our calculations reproduce the final hadron spectra as well,
at least, as those other works do. As a matter of fact, by using
$\tau_0$ = 0.15 fm/$c$ (rather than 0.6 fm/$c$), the system has a
few more tenths of fm/$c$ to develop some extra transverse collective
flow and there is no need to consider in our initial conditions a
supplemental input radial flow velocity parameter, $v_{r_0}$, as done 
in other works~\cite{peressou,kolb_rapp_flow} in order to reproduce the hadron 
spectra.

\subsection{Comparison to hadron data}

\begin{figure*}[htbp]
\psfig{figure=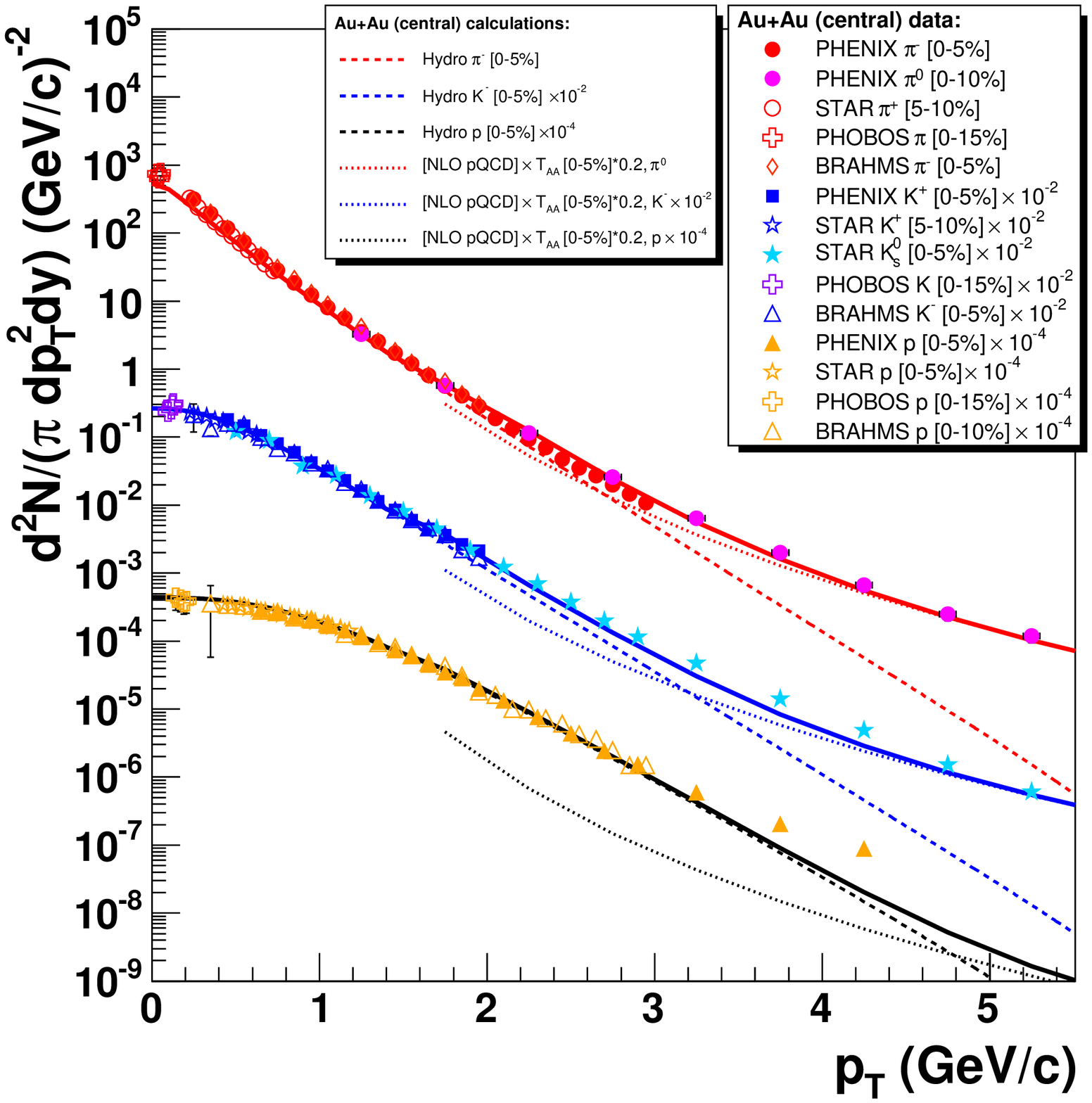,width=8cm,height=9cm}
\psfig{figure=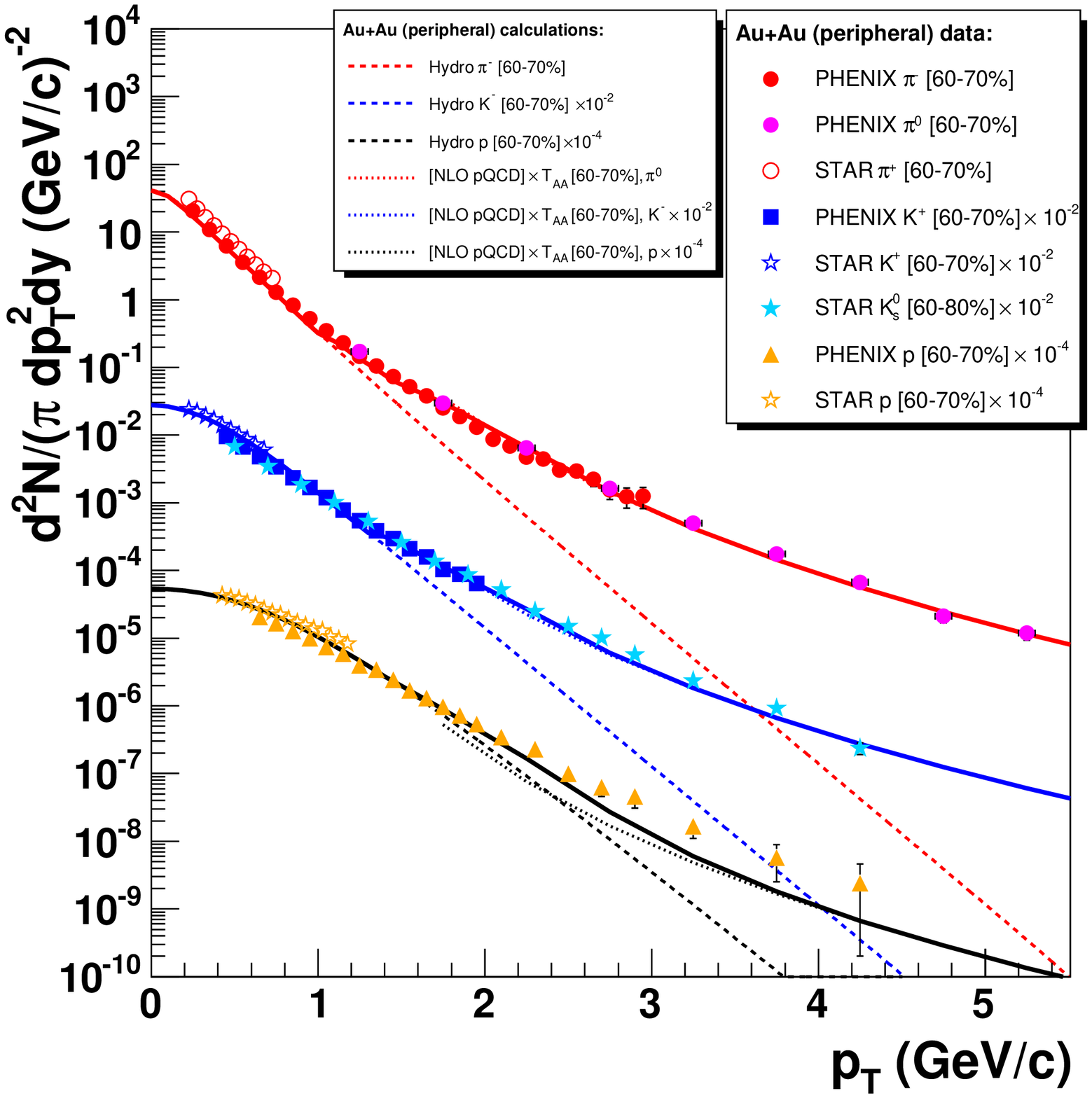,width=8cm,height=9cm}
\caption{Transverse momentum spectra for $\pi^{\pm,0}$,$K^{\pm,0}$, and protons
measured in the range $p_T$ = 0 -- 5.5 GeV/$c$ by PHENIX~\protect\cite{ppg026},
STAR ($K^0_s$ are preliminary)~\protect\cite{star_hadrons,star_hadrons2},
PHOBOS~\protect\cite{phobos_lowpt_had} and
BRAHMS~\protect\cite{brahms_hadrons} in central (0-10\% centrality, left)
and peripheral (60-70\%, right) Au+Au collisions at
$\sqrt{s_{\mbox{\tiny{\it{NN}}}}}$ = 200 GeV, compared to our hydrodynamics 
calculations (dashed lines), to the scaled pQCD p+p rates~\protect\cite{vogel_hadrons} 
(dotted lines), and to the sum hydro+pQCD (solid lines).}
\label{fig:hadron_spectra_AuAu200GeV}
\end{figure*}

Figure~\ref{fig:hadron_spectra_AuAu200GeV} shows the pion, kaon,
and proton\footnote{For a suitable comparison to the (feed-down
corrected) PHENIX~\cite{ppg026}, PHOBOS~\cite{phobos_lowpt_had}
and BRAHMS~\cite{brahms_hadrons} yields, the STAR proton
spectra~\cite{star_hadrons} have been appropriately corrected for
a $\sim$40\% ($p_T$-independent) contribution from weak
decays~\cite{star_hadrons2}.} transverse spectra measured by
PHENIX~\cite{ppg026}, STAR~\cite{star_hadrons,star_hadrons2},
PHOBOS~\cite{phobos_lowpt_had} and
BRAHMS~\cite{brahms_hadrons} in central (0--10\% corresponding
to $\mean{b}$ = 2.3 fm) and peripheral (60--70\% corresponding to $\mean{b}$ = 11.9 fm)
Au+Au collisions at $\sqrt{s_{\mbox{\tiny{\it{NN}}}}}$ = 200 GeV,
compared to our hydrodynamical predictions (dashed lines)
and to properly scaled p+p NLO pQCD expectations~\cite{vogel_hadrons}
(dotted lines). At low transverse momentum, the agreement data--hydro
is excellent 
starting from the very low $p_T$ PHOBOS data ($p_T<$ 100 MeV)
up to at least $p_T\approx$ 1.5 GeV/$c$.
Above this value, contributions from perturbative
processes (parton fragmentation products) start to dominate
over bulk hydrodynamic production.
Indeed, particles with transverse momenta $p_T\gtrsim$ 2 GeV/$c$
are mostly produced in primary parton-parton collisions at times
of order $\tau\sim 1/p_T\lesssim$ 0.15 fm/$c$ (i.e. during the
interpenetration of the colliding nuclei and {\it before}
any sensible time estimate for equilibration),
and as such, they are {\it not} in thermal equilibrium with the
bulk particle production. Therefore, one does not expect
hydrodynamics to reproduce the spectral shapes beyond $p_T\approx$ 2 GeV/$c$.
The dotted lines of Fig.~\ref{fig:hadron_spectra_AuAu200GeV} show
NLO predictions for $\pi$, $K$ and $p$ production in p+p
collisions at $\sqrt{s}$ = 200 GeV~\cite{vogel_hadrons} scaled by
the number of point-like collisions ($N_{coll}\propto T_{AA}$) times
an empirical quenching factor, $R_{AA}$ = 0.2 (0.7) for 0-10\% central
(60-70\% peripheral) Au+Au, to account for the observed constant suppression
factor of hadron yields at high $p_T$~\cite{phenix_hiptpi0_200,star_hipt_200}
(such a suppression is not actually observed in the $p,\bar{p}$ spectra 
at intermediate $p_T\approx$ 3 -- 5 GeV/$c$, see discussion below).\\

\begin{figure*}[htbp]
\psfig{figure=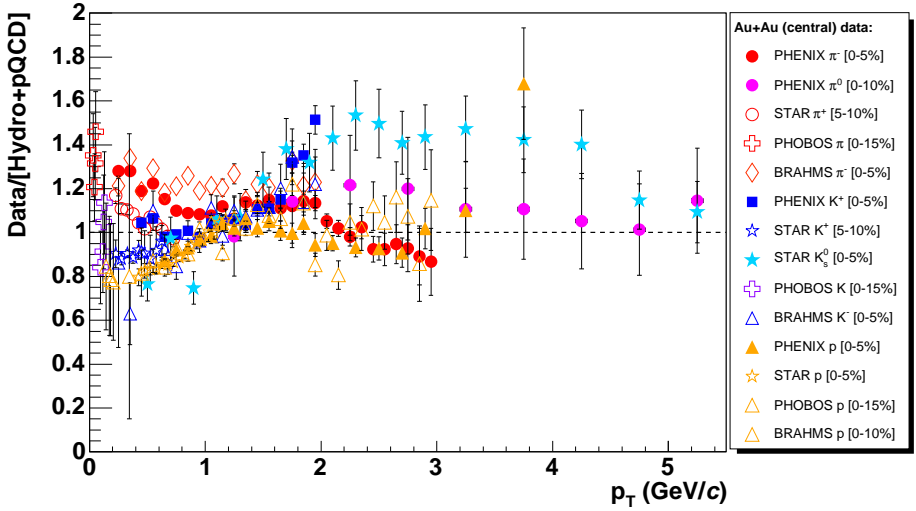,height=7.cm}
\caption{Ratio of $\pi^{\pm,0}$,$K^{\pm,0}$, and protons yields 
measured in the range $p_T$ = 0 -- 5.5 GeV/$c$ by PHENIX~\protect\cite{ppg026},
STAR (note that $K^0_s$ are preliminary)~\protect\cite{star_hadrons,star_hadrons2},
PHOBOS~\protect\cite{phobos_lowpt_had} and BRAHMS~\protect\cite{brahms_hadrons} 
in 0-10\% most central  Au+Au collisions at $\sqrt{s_{\mbox{\tiny{\it{NN}}}}}$ = 200 GeV, 
over the sum hydro ($\mean{b}$ = 2.3 fm) plus (quenched) pQCD. Theoretical calculations 
above $p_T\approx$ 2 GeV/$c$ have an overall $\pm$20\% uncertainty (not shown) dominated
by pQCD scale uncertainties.}
\label{fig:ratios_hadron_spectra_theory}
\end{figure*}

Fig.~\ref{fig:ratios_hadron_spectra_theory} shows more clearly (in linear rather than 
log scale as the previous figure) the relative agreement between the experimental hadron
transverse spectra and the hydrodynamical plus (quenched) pQCD yields presented in this work. 
The data-over-theory ratio plotted in the figure is obtained by taking the quotient of 
the pion, kaon and proton data measured in central Au+Au reactions (shown in the left plot 
of Fig.~\ref{fig:hadron_spectra_AuAu200GeV}) over the corresponding sum of hydrodynamical 
plus perturbative results (solid lines in Fig.~\ref{fig:hadron_spectra_AuAu200GeV}).
In the low $p_T$ range dominated by hydrodynamical production, there exist some local 
$p_T$-dependent deviations between the measurements and the calculations. However, 
the same is true within the independent data sets themselves and, thus, those differences
are indicative of the amount of systematic uncertainties associated with the different 
measurements. High $p_T$ hadro-production, dominated by perturbative processes, agrees
also well within the $\sim$20\% errors associated with the standard scale uncertainties
for pQCD calculations at this center-of-mass energy. 
It is, thus, clear from Figs.~\ref{fig:hadron_spectra_AuAu200GeV} 
and~\ref{fig:ratios_hadron_spectra_theory} that
identified particle production at $y$ = 0 in nucleus-nucleus
collisions at RHIC can be fully described in their whole $p_T$ range
and for all centralities by a combination of hydrodynamical
(thermal+collective boosted) emission plus (quenched) prompt
perturbative production. An exception to this rule are, however, 
the (anti)protons~\cite{phnx_ppbar}. Although due to their higher masses,
they get an extra push from the hydrodynamic flow up to
$p_T\sim 3 $ GeV/$c$, for even higher transverse momenta the
combination of hydro plus (quenched) pQCD still clearly undershoots the
experimental proton spectra. This observation has lent support to the 
existence of an additional mechanism for baryon production at intermediate 
$p_T$ values ($p_T\approx$ 3 -- 5 GeV/$c$)
based on quark recombination~\cite{recomb}. This mechanism will not,
however, be further considered in this paper since it has no practical
implication for photon production and/or for the overall hydrodynamical
evolution of the reaction. The overall good theoretical reproduction of the
differential $\pi,K,p$ experimental spectra for all centralities is obviously
consistent with the previous observation that our calculated total integrated hadron 
multiplicities agree very well with the experimental data measured at
mid-rapidity by the four different RHIC experiments (Fig.~\ref{fig:dNch}).

\section{Direct photon production}

As in the case of hadron production, the total direct photon
spectrum in a given Au+Au collision at impact parameter $b$ is
obtained by adding the primary production from perturbative
parton-parton scatterings to the thermal emission rates integrated
over the whole space-time volume of the produced fireball. Three
sources of direct photons are considered
corresponding to each one of the phases of the reaction: prompt
production, partonic gas emission, and hadronic gas radiation.

\subsection{Prompt photons}
 
\begin{sloppypar}
For the prompt $\gamma$ production we use the NLO pQCD predictions of
W.~Vogelsang~\cite{vogel_gamma} scaled by the corresponding
Glauber nuclear overlap function at $b$, $T_{AA}(b)$, as expected
for hard processes in A+A collisions unaffected by final-state effects 
(as empirically confirmed for photon production in Au+Au~\cite{ppg042}).
This pQCD photon spectrum is obtained with CTEQ6M~\cite{cteq6} parton 
distribution function (PDF), GRV~\cite{grv_photons} parametrization of 
the $q,g\rightarrow\gamma$ fragmentation function (FF), and
renormalization-factorization scales set equal to the transverse
momentum of the photon ($\mu = p_T$). Such NLO calculations
provide an excellent reproduction of the inclusive direct 
$\gamma$~\cite{ppg049} and large-$p_T$ $\pi^0$~\cite{ppg024}
spectra measured by PHENIX in p+p
collisions at $\sqrt{s}$ = 200 GeV without any additional
parameter (in particular, at variance with results at lower
energies~\cite{wong}, no primordial $k_T$ is needed to describe
the data). We do {\it not} consider any modification of the prompt
photon yields in Au+Au collisions due to partially counteracting
initial-state (IS) effects such as: (i) nuclear modifications
(``shadowing'') of the Au PDF 
($<20$\%, in the relevant ($x,Q^2$) kinematical range considered
here~\cite{frankfurt_rhic_lhc,ina,jamal}), and (ii) extra nuclear
$k_T$ broadening (Cronin enhancement) as described e.g. in~\cite{dumitru}.
Both IS effects are small and/or approximately cancel each other at mid-rapidity 
at RHIC as evidenced experimentally by the barely modified nuclear modification 
factor, $R_{dAu}\lesssim$1.1, for $\gamma$ and $\pi^0$ measured in d+Au collisions 
at $\sqrt{s_{\mbox{\tiny{\it{NN}}}}}$ = 200 GeV~\cite{qm05}.
Likewise, we do {\it not} take into account any
possible final-state (FS) {\it photon} suppression due to energy
loss of the jet-fragmentation (aka. ``anomalous'') component of the
prompt photon cross-section~\cite{dumitru,jamal,arleo04,dde_hq04}, which, 
if effectively present (see~\cite{zakharov04} and discussion in Sect.~\ref{sec:extra_gamma}), 
can be in principle experimentally determined by detailed measurements of the 
isolated and non-isolated direct photon baseline spectra in p+p collisions
at $\sqrt{s}$ = 200 GeV~\cite{dde_hq04}.
\end{sloppypar}

\subsection{Thermal photon rates}

\begin{sloppypar}
For the QGP phase we use the most recent full leading order (in $\alpha_{em}$ 
and $\alpha_s$ couplings) emission rates from Arnold {\it et al.}~\cite{arnold}.
These calculations include hard thermal loop diagrams to all orders and
Landau-Migdal-Pomeranchuk (LPM) medium interference effects.
The parametrization given in~\cite{arnold}
assumes zero net baryon density (i.e. null quark chemical
potential, $\mu_q$ = 0), and {\it chemical} together with thermal
equilibrium. Corrections of the QGP photon rates due to net quark
densities are $\mathcal{O}[\mu_q^2/(\pi T)^2]$~\cite{traxler} i.e.
marginal at RHIC energies where the baryochemical potential is
close to zero at midrapidity ($\mu_B=3\,\mu_q\sim$ 25 MeV) and
neglected here. Similarly, although the early partonic phase is
certainly not chemically equilibrated (the first instants of the
reaction are strongly gluon-dominated) the two main effects from
chemical non-equilibrium composition of the QGP: reduction of
quark number and increase of the temperature, nearly cancel in the
photon spectrum~\cite{yellow_rep,gelis} and have not been
considered either.
For the HRG phase, we use the latest improved parametrization
from Turbide {\it et al.}~\cite{turbide} which includes hadronic
emission processes 
not accounted for before in the literature. In all calculations,
we use a temperature-dependent 
parametrization of the strong coupling\footnote{According to this parametrization,
$\alpha_S(T)$ = 0.3 -- 0.6 in the range of temperatures of interest here ($T\approx$ 600 -- 150 MeV).},
$\alpha_s(T) = 2.095/\{\frac{11}{2\pi}\ln{(Q/\Lambda_{\overline{MS}})} +
\frac{51}{22\pi}\ln{[2\ln(Q/\Lambda_{\overline{MS}})]}\}$ with $Q = 2\pi T$,
obtained from recent lattice results~\cite{karsch_alphaS}.
\end{sloppypar}

\subsection{Extra photon contributions} 
\label{sec:extra_gamma}

Apart from the aforementioned (prompt and thermal) photon production mechanisms,
S.~Bass {\it et al.}~\cite{bass,bass2} have recently evaluated within
the Parton Cascade Model (PCM), the contribution to the total
Au+Au photon spectrum from secondary (cascading) parton-parton
collisions taking place before the attainment of thermalization
(i.e. between the transit time of the two nuclei, $\tau \approx$ 0.15 fm/$c$,
and the standard $\tau_{therm}$ = 0.6 fm/$c$ considered at RHIC).
Since such cascading light emission is due to second-chance
partonic collisions which are, simultaneously, driving the system
towards equilibrium, we consider not only ``valid'' (see the discussion of
refs.~\cite{berges04,arnold04}) but more self-consistent within our framework 
to account for this contribution 
with our hydrodynamical evolution alone. We achieve this by starting 
hydrodynamics (whose photon rates also include the expected LPM reduction of 
the secondary rates~\cite{bass2}) at $\tau_{0}$ = 0.15 fm/$c$. By doing that, 
at the same time that we account for this second-chance emission, our initial 
plasma temperature and associated thermal photon production can be
considered to be at their {\it maximum values} for RHIC energies.\\ 

\begin{figure*}[htbp]
\psfig{figure=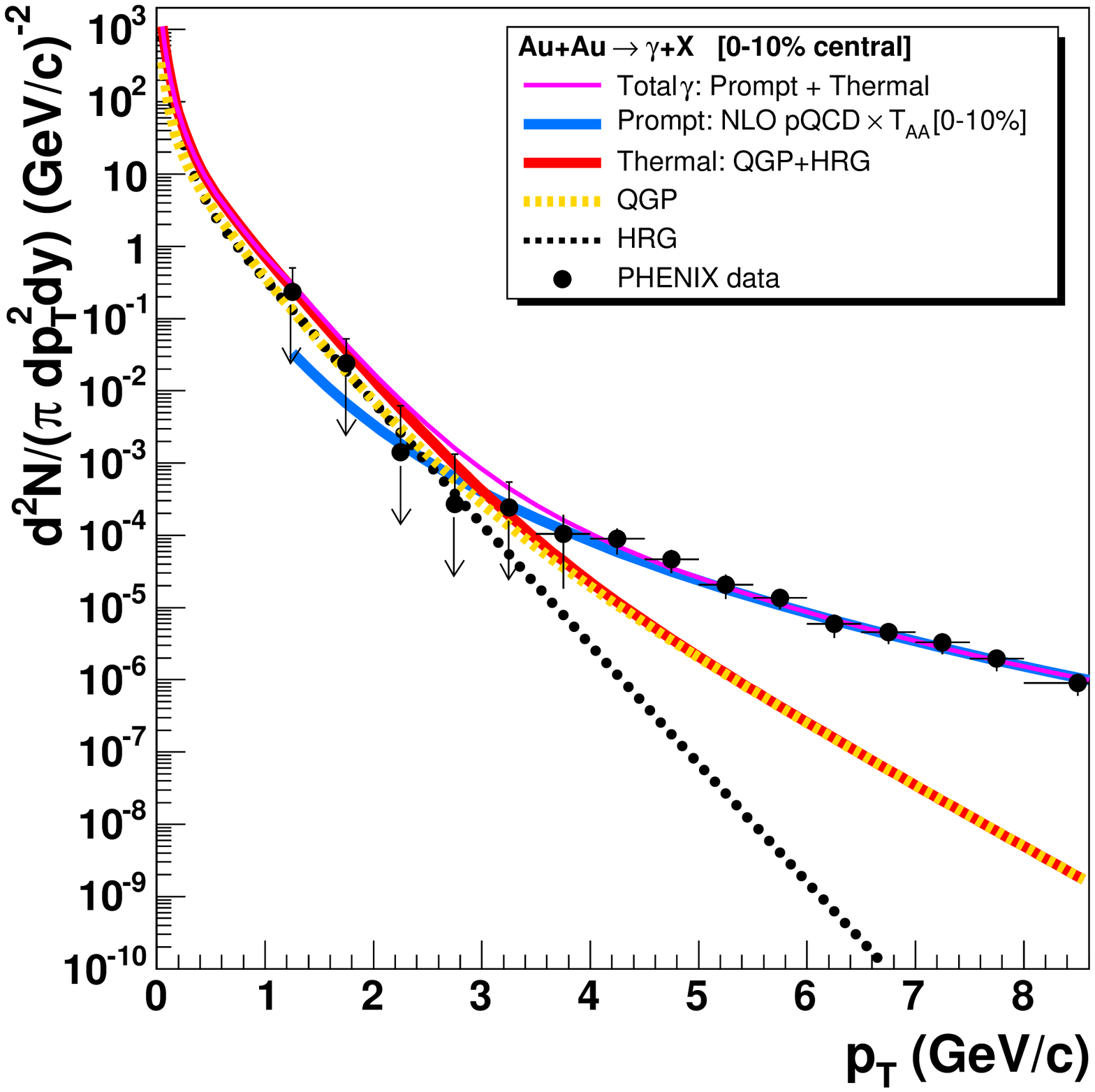,height=9.cm}
\psfig{figure=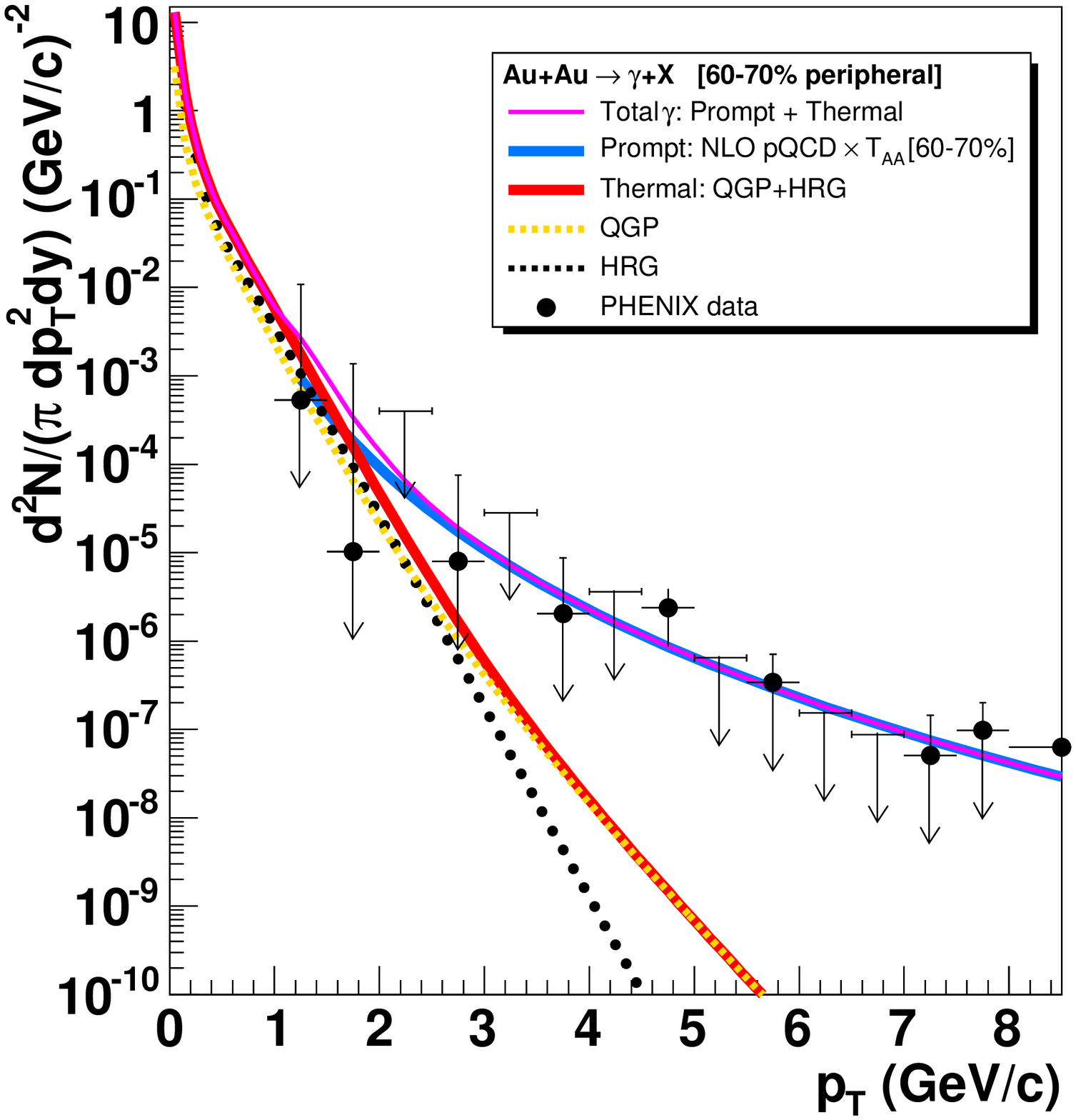,height=9.cm}
\caption{Photon spectra for central (0--10\%, left) and peripheral
(60--70\%, right) Au+Au reactions at $\sqrt{s_{\mbox{\tiny{\it{NN}}}}}$ = 200 GeV
as computed with our hydrodynamical model [with the contributions
for the QGP and hadron resonance gas (HRG) given separately]
compared to the expected NLO pQCD p+p yields for the prompt
$\gamma$~\protect\cite{vogel_gamma} (scaled by the corresponding
nuclear overlap function), and to the experimental photon
yields measured by the PHENIX collaboration~\protect\cite{ppg042}.}
\label{fig:photon_spec_AuAu_cent_periph}
\end{figure*}

Likewise, we do not consider the conjectured extra $\gamma$ emission 
due to the passage of quark jets (Compton-scattering and annihilating) through 
the dense medium~\cite{fries_rhic_lhc,fries_gammajet2,zakharov04} since such contribution 
is likely partially compensated by: (i) the concurrent non-Abelian 
energy loss of the parent quarks going through the system~\cite{turbide05},
plus (ii) a possible {\it photon} suppression due to energy loss of the 
``anomalous'' component of the prompt photon cross-section~\cite{dumitru,jamal,dde_hq04,arleo04}.
As a matter of fact, some approximate cancellation of all those effects must exist
since the experimental Au+Au photon spectra above $p_T\approx$ 4 GeV/$c$ turn out to be
well reproduced by primary (pQCD) hard processes alone for all centralities, 
as can be seen in the comparison of pQCD NLO predictions with PHENIX data~\cite{ppg042} 
(Fig.~\ref{fig:photon_spec_AuAu_cent_periph}). The apparent agreement between the
experimental spectra above $p_T\approx$ 4 GeV/$c$ and the NLO calculations does not
seem to leave much room for extra radiation contributions. A definite conclusion on the
existence or not of FS effects on photon production will require in any case
precision $\gamma$ data in Au+Au, d+Au and p+p collisions. The more critical
issue of the role of the jet bremsstrahlung component needs to be estimated, for
example, via measurements of isolated and non-isolated direct photon baseline 
p+p spectra as discussed in~\cite{dde_hq04}. Additional IS effects  not considered so far
due, for example, to isospin corrections\footnote{Direct photon cross-sections 
depend on the light quark electric charges and are thus disfavoured in a nucleus target
less rich in up quarks than the standard proton reference~\cite{arleo05}.} will require a 
careful analysis and comparison of Au+Au to reference d+Au photon cross-sections too.\\

\subsection{Total direct photon spectra} 

\begin{figure}[htbp]
\psfig{figure=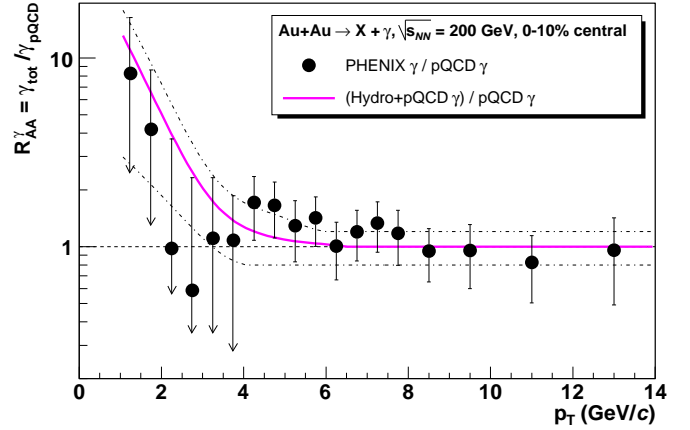,height=6.cm}
\caption{Direct photon ``nuclear modification factor'', $R_{AA}^\gamma$ (Eq.~\ref{eq:RAA}), 
obtained as the ratio of the total over the prompt $\gamma$ spectra for 
0--10\% most central Au+Au reactions at $\sqrt{s_{\mbox{\tiny{\it{NN}}}}}$ = 200 GeV. 
The solid line is the ratio resulting from our hydro+pQCD model.
The points show the PHENIX data~\protect\cite{ppg042} over the same NLO yields and
the dashed-dotted curves indicate the theoretical uncertainty of the NLO calculations
(see text).}
\label{fig:RAA_photon}
\end{figure}

Figure~\ref{fig:photon_spec_AuAu_cent_periph} shows our computed 
total direct photon spectra for central (left) and peripheral (right)
Au+Au collisions at $\sqrt{s_{\mbox{\tiny{\it{NN}}}}}$ = 200 GeV,
with the pQCD, QGP, and HRG components differentiated\footnote{We
split the mixed phase contribution onto QGP and HRG components
calculating the relative proportion of QGP (HRG) matter in it.}.
In central reactions, thermal photon production (mainly of QGP origin)
outshines the prompt pQCD emission below $p_T\approx$ 3 GeV/$c$.
Within $p_T\approx$ 1 -- 4 GeV/$c$, thermal photons account for
roughly 90\% -- 50\% of the total photon yield in central Au+Au, 
as can be better seen in the ratio total-$\gamma$/pQCD-$\gamma$
shown in Fig.~\ref{fig:RAA_photon}. Photon production in peripheral 
collisions is, however, clearly dominated by the primary parton-parton radiation.
In both cases, hadronic gas emission prevails only for lower $p_T$ values.
In Figure~\ref{fig:photon_spec_AuAu_cent_periph} we also compare our
computed spectra to the inclusive Au+Au photon spectra published
recently by the PHENIX collaboration~\cite{ppg042}.
The total theoretical (pQCD+hydro) differential cross-sections are in
good agreement with the experimental yields, though for central reactions 
our calculations tend to ``saturate'' the upper limits of the data 
in the range below $p_T\approx$ 4 GeV/$c$ where thermal photons dominate. 
New preliminary PHENIX Au+Au direct-$\gamma^*$ results~\cite{qm05,qm05_akiba} 
are also systematically above (tough still consistent with) these published 
spectra in the range $p_T\approx$ 1 -- 4 GeV/$c$ and, if confirmed, will 
bring our results to an even better agreement with the data.\\

To better distinguish the relative amount of thermal radiation in the theoretical 
and experimental {\it total} direct photon spectra in central Au+Au collisions, 
we present in Figure~\ref{fig:RAA_photon} the nuclear modification factor 
$R_{AA}^\gamma$ defined as the ratio of the total over prompt (i.e. $T_{AA}$-scaled 
p+p pQCD predictions) photon yields: 
\begin{equation}
R_{AA}^{\gamma}(p_T)\;=\;\frac{dN_{AuAu}^{total\;\gamma}/dp_{T}}{T_{AA}\cdot d\sigma_{pp}^{\gamma\;pQCD}/dp_{T}}.
\label{eq:RAA}
\end{equation}

A value $R_{AA}^\gamma \approx$ 1 would indicate that all the photon yield can be 
accounted for by the prompt production alone.
Of course, since our total direct-$\gamma$ result for central Au+Au includes thermal 
emission from the QGP and HRG phases, we theoretically obtain $R_{AA}^\gamma \approx$ 10 -- 1 
in the $p_T\approx$ 1 -- 4 GeV/$c$ region where the thermal component is significant 
(Fig.~\ref{fig:RAA_photon}). In this very same $p_T$ range, although the available PHENIX 
results have still large uncertainties\footnote{Technically, the PHENIX data points below 
$p_T$ = 4 GeV/$c$ have ``lower errors that extend to zero'', i.e. a non-zero direct-$\gamma$ signal is 
indeed observed in the data but the associated errors are larger than the signal itself~\cite{ppg042}.}, 
the central value of most of the data points is clearly consistent with the existence of 
a significant excess over the NLO pQCD expectations. A note of caution is worth here, however, 
regarding the $R_{AA}^\gamma\gg$ 1 value observed for both the theoretical and experimental 
spectra below $p_T\approx$ 4 GeV/$c$ since it is not yet clear to what extent the 
NLO predictions, entering in the denominator of Eq.~(\ref{eq:RAA}), are realistic in 
this thermal-photon ``region of interest''. 
Indeed, in this comparatively low $p_T$ range the theoretical prompt yields are dominated 
by the jet bremsstrahlung contribution~\cite{dde_hq04} which is intrinsically non-perturbative 
(i.e. not computable) and determined solely from the parametrized parton-to-photon 
GRV~\cite{grv_photons} FF which is relatively poorly known in this kinematic range. 
The standard scale uncertainties in the NLO pQCD calculations are $\pm$20\% above
$p_T\approx$ 4 GeV/$c$ but we have assigned a much more pessimistic $_{+50}^{-200}$\%
uncertainty to these calculations in the range $p_T\approx$ 1 -- 4 GeV/$c$ (dashed-dotted
lines in Fig.~\ref{fig:RAA_photon}). Precise measurements of the direct-$\gamma$ baseline 
spectrum in p+p collisions at $\sqrt{s}$ = 200 GeV above $p_T$ = 1 GeV/$c$ are mandatory 
before any definite conclusion can be drawn on the existence or not of a thermal excess
from the Au+Au experimental data.\\


\begin{figure}[htbp]
\begin{center}
\psfig{figure=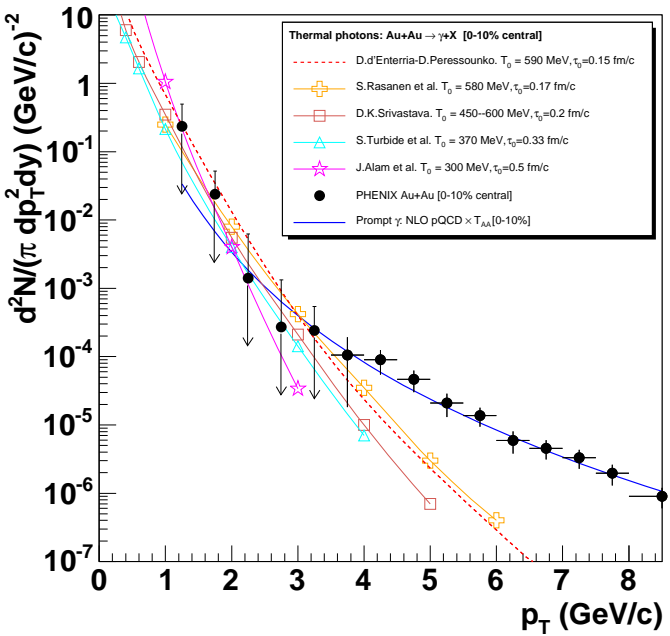,height=8.5cm}
\end{center}
\caption{Thermal photon predictions for central Au+Au reactions at $\sqrtsnn$ = 200 GeV 
as computed with different hydrodynamical~\protect\cite{srivastava_sps_rhic,alam_sps_rhic,finnish_hydro}
or ``dynamical fireball''~\protect\cite{turbide} models, compared to (i) our hydro calculations 
(dashed curve), (ii) the expected perturbative $\gamma$ yields ($T_{AA}$-scaled 
NLO p+p calculations~\protect\cite{vogel_gamma}), and (iii) the experimental total direct photon spectrum 
measured by PHENIX~\protect\cite{ppg042}.}
\label{fig:compare}
\end{figure}

As a final cross-check of our computed hydrodynamical photon yields, we have compared
them to previously published predictions for thermal photon production in Au+Au collisions 
at top RHIC energy: D.~K.~Srivastava {\it et al.}~\cite{srivastava_sps_rhic}
(with initial conditions $\tau_0\approx$ 0.2 fm/$c$ and $T_0\approx$ 450 -- 660 MeV), 
Jan-e Alam {\it et al.}\footnote{Alam {\it et al.} have recently~\cite{alam05} recomputed their 
hydrodynamical yields using higher initial temperatures ($T_0$ = 400 MeV 
at $\tau_0$ = 0.2 fm/$c$) and getting a better agreement with the data.}~\cite{alam_sps_rhic} 
($\tau_0$ = 0.5 fm/$c$ and $T_0$ = 300 MeV), 
F.~D.~Steffen and M.~H.~Thoma~\cite{steffen_sps_rhic_lhc} ($\tau_0$ = 0.5 fm/$c$ and
$T_0$ = 300 MeV), S.~S.~Rasanen {\it et al.}~\cite{finnish_hydro}
($\tau_0$ = 0.17 fm/$c$ and $T_0$ = 580 MeV), N.~Hammon {\it et al.}~\cite{frankfurt_rhic_lhc} 
($\tau_0$ = 0.12 fm/$c$ and $T_0$ = 533 MeV), and Turbide {\it et al.}\footnote{Note that
{\it stricto senso} Turbide's spectra are not obtained with a pure hydrodynamical computation but 
using a simpler ``dynamical fireball'' model which assumes constant acceleration in longitudinal 
and transverse directions.}~\cite{turbide} ($\tau_0$ = 0.33 fm/$c$ and $T_0$ = 370 MeV).
For similar initial conditions, the computed total thermal yields in those works are compatible 
within a factor of $\sim$2 with those presented here. Some of those predictions are shown 
in Figure~\ref{fig:compare} confronted to our calculations. Our yields are, in general,
above all other predictions since, as aforementioned, both our initial thermalization time
and energy densities (temperatures) have the most ``extreme'' values possible consistent with 
the RHIC charged hadron multiplicities. They agree specially well with the hydrodynamical 
calculations of the Jyv\"askyl\"a group~\cite{finnish_hydro} which have been computed 
with the same up-to-date QGP rates used here. 
Given the current (large) uncertainties of the available published data, all thermal photon 
predictions are consistent with the experimental results. However, as aforementioned, newer 
(preliminary) PHENIX direct-$\gamma^\star$ measurements 
have been reported very recently~\cite{qm05,qm05_akiba} and indicate a clear excess of direct photons 
over NLO pQCD for Au +Au at $\sqrtsnn$ = 200 GeV in this $p_T$ range in excellent agreement 
with our thermal photon calculations.

\section{Thermal photons and the QCD equation-of-state}
\label{sec:eos}

In order to experimentally isolate the thermal photon spectrum one needs
to subtract from the total direct $\gamma$ spectrum the non-equilibrated
``background'' of prompt photons. The prompt $\gamma$ contribution emitted
in a given Au+Au centrality can be measured separately in
reference p+p (or d+Au) collisions at the same $\sqrt{s}$, scaled by the
corresponding nuclear overlap function $T_{AA}(b)$, and subtracted from the
total Au+Au $\gamma$ spectrum~\cite{dde_hq04}. The simpler expectation is
that the remaining photon spectrum for a given impact parameter $b$
\begin{equation}
\frac{dN_{AuAu}^{thermal\;\gamma}(b)}{dp_T} \; = \; \frac{dN_{AuAu}^{total\;\gamma}(b)}{dp_T}
- T_{AA}(b)\cdot\frac{d\sigma_{pp}^{\gamma}}{dp_T}\;,
\label{eq:thermal_spec}
\end{equation}
will be just that due to thermal emission from the partonic and hadronic phases 
of the reaction. Such a subtraction procedure can be effectively applied to all the
$\gamma$ spectra measured in different centralities as long as both the total Au+Au 
and baseline p+p photon spectra are experimentally measured with reasonable 
($\lesssim$15\%) point-to-point (systematical and statistical) uncertainties~\cite{dde_hq04}.
The subtracted spectra~(\ref{eq:thermal_spec}) can be therefore subject to 
scrutiny in terms of the thermodynamical properties of the radiating medium.

\subsection{Determination of the initial temperature}

\begin{sloppypar}
Due to their weak electromagnetic interaction with the surrounding medium, 
photons produced in the reaction escape freely the interaction region immediately 
after their production. Thus, even when emitted from an equilibrated source, they are 
not reabsorbed by the medium and do not have a black-body spectrum at the source temperature. 
Nonetheless, since all the theoretical thermal $\gamma$ rates~\cite{arnold,turbide}
have a general functional dependence of the form\footnote{The $T^2$ factor is 
just an overall normalization factor in this case (since its temporal variation is 
small compared to the short emission times) and does not significantly alter the 
exponential shape of the spectra.} $E_\gamma\,dR_{\gamma}/d^3p\propto T^2\cdot \exp{(-E_{\gamma}/T)}$,
one would expect the final spectrum to be locally exponential with an inverse slope 
parameter strongly correlated with the (local) temperature $T$ of the radiating medium.
Obviously, such a general assumption is complicated by several facts.
On the one hand, 
the final thermal photon spectrum is a sum of exponentials with different temperatures 
resulting from emissions at different time-scales and/or from different
regions of the fireball which has strong temperature gradients (the core being much hotter 
than the ``periphery''). On the other hand, collective flow effects (stronger for 
increasingly central collisions) superimpose on top of the purely thermal emission 
leading to an effectively larger inverse slope parameter ($T_{\ensuremath{\it
eff}}\approx\sqrt{(1+\beta)/(1-\beta)}\,T$)~\cite{peressou}.
One of the main results of this paper is to show that, based
upon a realistic hydrodynamical model, such effects do not 
destroy completely the correlation between the apparent photon temperature
and the maximal temperature actually reached at the beginning of 
the collision process. We will show that such a correlation indeed
exists and that the local inverse slope parameter obtained by fitting 
to an exponential, at high enough $p_T$, the thermal photon spectrum obtained 
via the expression (\ref{eq:thermal_spec}), indeed provides a good proxy of the 
initial temperature of the system without much 
distortion due to collective flow (and other) effects.\\
\end{sloppypar}

\begin{figure}[htbp]
\centerline{\psfig{figure=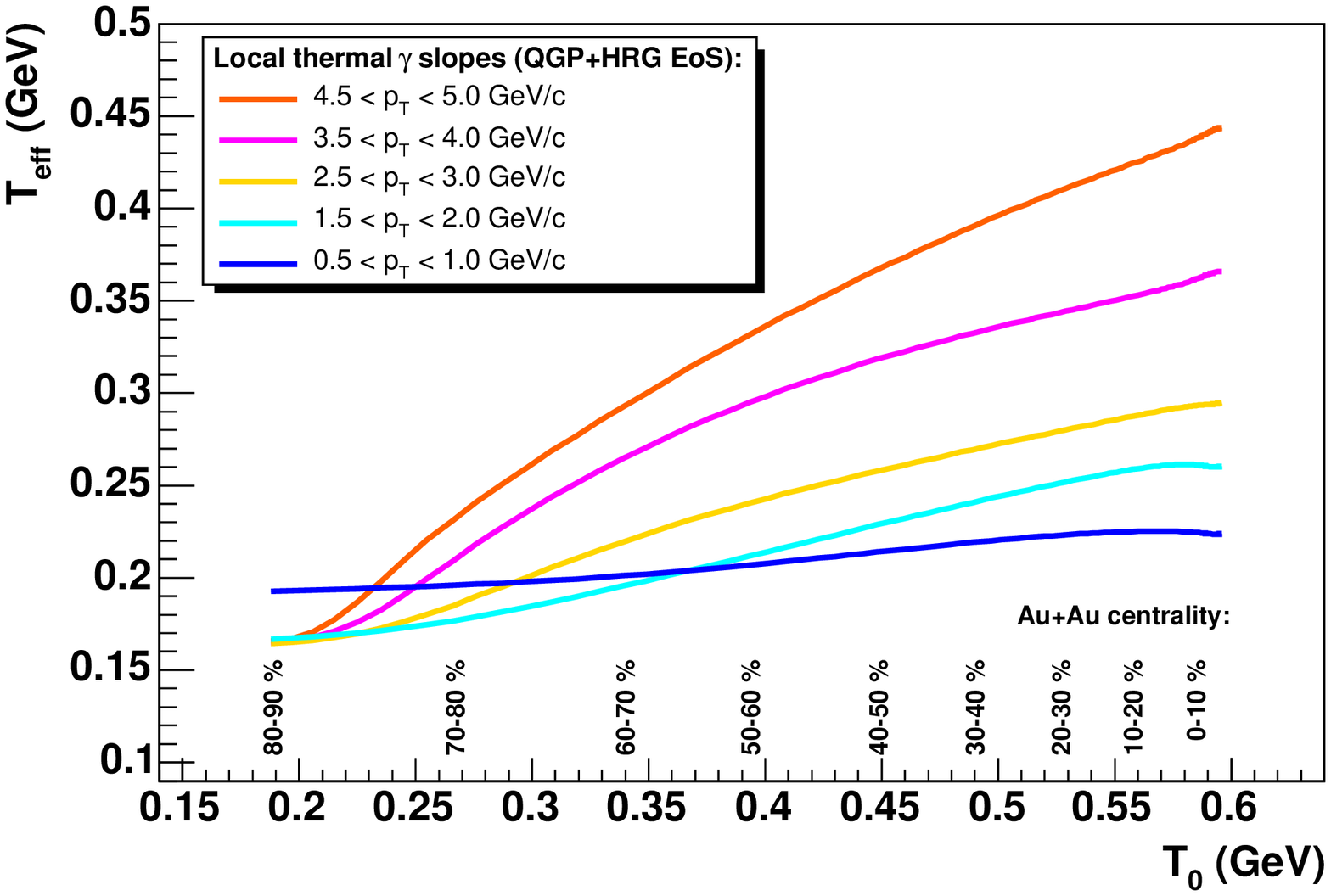,height=6.cm,width=9.5cm}}
\centerline{\psfig{figure=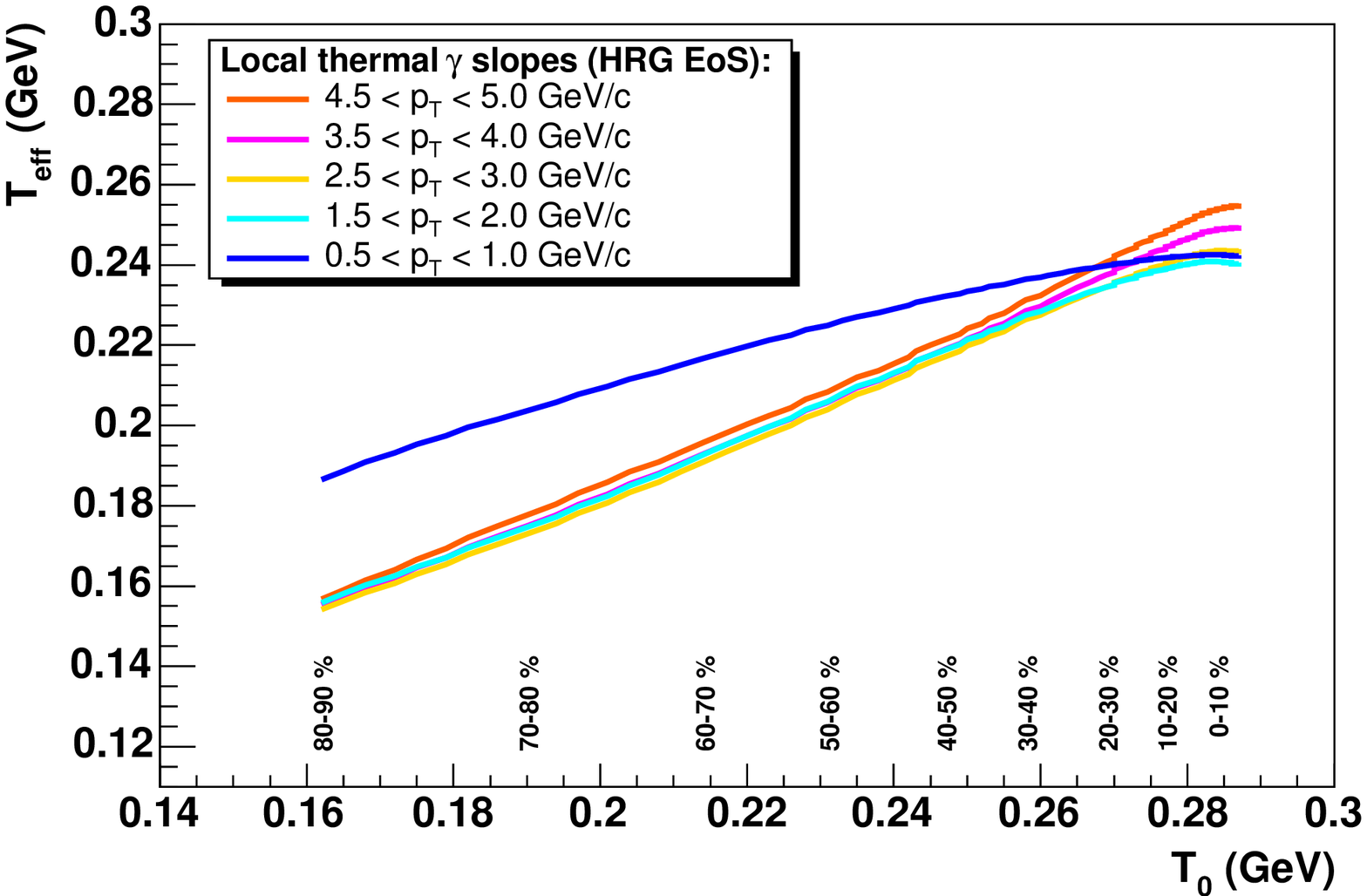,height=5.cm,width=9.5cm}}
\caption{Local photon slope parameters $T_{\ensuremath{\it eff}}$ 
(obtained from exponential fits of the thermal photon
spectrum in different $p_T$ ranges) plotted versus the initial
(maximum) temperature $T_{0}$ of the fireball produced at different 
centralities in Au+Au collisions at $\sqrt{s}=200$ GeV. Upper
plot - hydrodynamical calculations with QGP+HRG EoS
(Table~\ref{tab:hydro_parameters}), bottom - HRG EoS (with initial
conditions: $\varepsilon_0 = 30$ GeV/fm$^3$ at
$\tau_0=0.6$~fm/$c$).}
\label{fig:thermal_slopes}
\end{figure}

To determine to what extent the thermal slopes are indicative of the original 
temperature of the system, we have fitted the thermal spectra obtained from our 
hydrodynamical calculations in different Au+Au centralities to an exponential 
distribution in different $p_T$ ranges. Since, -- according to our Glauber prescription 
for the impact-parameter dependence of the hydrodynamical initial conditions --,
different centralities result in different initial energy densities, we can
in this way explore the dependence of the apparent thermal photon temperature on 
the maximal initial temperatures $T_{0}$ (at the core) of the system. The upper plot 
of Figure~\ref{fig:thermal_slopes} shows the obtained local slope parameter, 
$T_{\ensuremath{\it eff}}$, as a function of the initial 
$T_{0}$ for our default QGP+HRG hydrodynamical evolution (Table~\ref{tab:hydro_parameters}). 
We find that although all the aforementioned effects smear the correlation between 
the apparent and original temperatures, they do not destroy it completely.
The photon slopes are indeed approximately proportional to the initial temperature
of the medium, $T_{0}$. There is also an obvious anti-correlation between the $p_T$
of the radiated photons and their time of emission.
At high enough $p_T$ the hardest photons issuing from the
hottest zone of the system swamp completely any other softer contributions
emitted either at later stages and/or from outside the core region of the fireball.
Thus, the higher the $p_T$ range, the closer is $T_{\ensuremath{\it eff}}$ 
to the original $T_{0}$ at the center of the system. According to our calculations, 
empirical thermal slopes measured above $p_T\approx$ 4 GeV/$c$ in central Au+Au collisions 
are above $\sim$400 MeV i.e. only $\sim$30\% lower than the ``true'' maximal (local) 
temperature of the quark-gluon phase. On the other hand, local $\gamma$ slopes 
in the range below $p_T\approx$ 1 GeV/$c$ have almost constant value 
$T_{\ensuremath{\it eff}}\sim$ 200 MeV (numerically close to $T_{crit}$)
for all centralities and are almost insensitive to the initial temperature of the 
hydrodynamical system but mainly specified by the exponential prefactors in the 
hadronic emission rates, plus collective boost effects.\\

To assess the dependence of the thermal photon spectra on the underlying EoS, 
we have rerun our hydro evolution 
with just the EoS of a hadron resonance gas. We choose now as initial conditions: 
$\varepsilon_0 = 30$ GeV/fm$^3$ at $\tau_0=0.6$~fm/$c$, 
which can still reasonably describe the experimental hadron spectra.
Obviously, any description in terms of hadronic degrees of freedom
at such high energy densities is unrealistic but we are interested
in assessing the effect on the thermal photon slopes of a non ideal-gas EoS 
as e.g. that of a HRG-like system with a large number of heavy resonances 
(or more generally, of any EoS with exponentially rising number of mass states).
The photon slopes for the pure HRG gas EoS (Fig.~\ref{fig:thermal_slopes}, bottom)
are lower ($T_{\ensuremath{\it eff}}^{\ensuremath{\it max}}\approx$ 220 MeV)
than in the default QGP+HRG evolution, not only because the input HRG $\varepsilon_0$
is smaller (the evolution starts at a later $\tau_0$) but, specially because for the 
same initial $\varepsilon_0$ the effective number of degrees of freedom in a system 
with a HRG EoS is higher than that in a QGP\footnote{Note that $g(T)\propto \varepsilon/T^4$ 
increases exponentially with $T$ for a HRG-like EoS, and at high enough temperatures will 
clearly overshoot the QGP constant number of degrees of freedom.}
and therefore the initial temperatures are lower. A second difference is that, for all 
$p_T$ ranges, we find almost the same exact correlation between the local $\gamma$ slope 
and $T_{0}$ indicating a single underlying (hadronic) radiation mechanism dominating 
the transverse spectra at all $p_T$.\\

Two overall conclusions can be obtained from the study of the hydrodynamical photon
slopes. First, the observation in the data, via Eq.~(\ref{eq:thermal_spec}), of a thermal 
photon excess  above $p_T\approx$ 2.5 GeV/$c$ with exponential slope 
$T_{\ensuremath{\it eff}}\gtrsim$ 250 MeV is an unequivocal proof of the formation 
of a system with maximum temperatures above $T_{crit}$ since no realistic collective 
flow mechanism can generate such a strong boost of the photon slopes, while simultaneously 
reproducing the hadron spectra.
Secondly, pronounced $p_T$ dependences of the local thermal slopes seem to be characteristic
of space-time evolutions of the reaction that include an ideal-gas QGP radiating phase.

\subsection{Determination of the QCD Equation of State (EoS)}

As we demonstrated in the previous section, $T_{\ensuremath{\it eff}}$ is 
approximately proportional to the maximum temperature reached in a nucleus-nucleus reaction. 
One can go one step further beyond the mere analysis of the thermal photon slopes 
and try to get a direct handle on the equation of state of the radiating medium by looking 
at the correlation of $T_{\ensuremath{\it eff}}$ with experimental observables 
related to the initial energy or entropy densities of the system.
For example, assuming an isentropic expansion (which is implicit in our perfect 
fluid hydrodynamical equations with zero viscosity) one can estimate the 
{\it initial} entropy density $s$ at the time of photon emission 
from the total {\it final} particle multiplicity $dN/dy$ measured in the reaction. 
Varying the centrality of the collision, one can then explore the form of the dependence $s\,=\,s(T)$ 
at the first instants of the reaction, extract the underlying equation of state
of the radiating system and trace any signal of a possible phase transition.
Indeed, the two most clear evidences of QGP formation from QCD calculations on
the lattice are: (i) the sharp rise of $\varepsilon(T)/T^4$, or equivalently
of $s(T)/T^3$, at temperatures around $T_{crit}$, and (ii) the flattening
of the same curve above $T_{crit}$. The sharp jump is of course due to the 
sudden release of a large number of (partonic) degrees of freedom at $T_{crit}$.
The subsequent plateau is due to the full formation of a QGP with a 
{\it fixed} (constant) number of degrees of freedom.\\

\begin{figure}[htbp]
\begin{center}
\psfig{figure=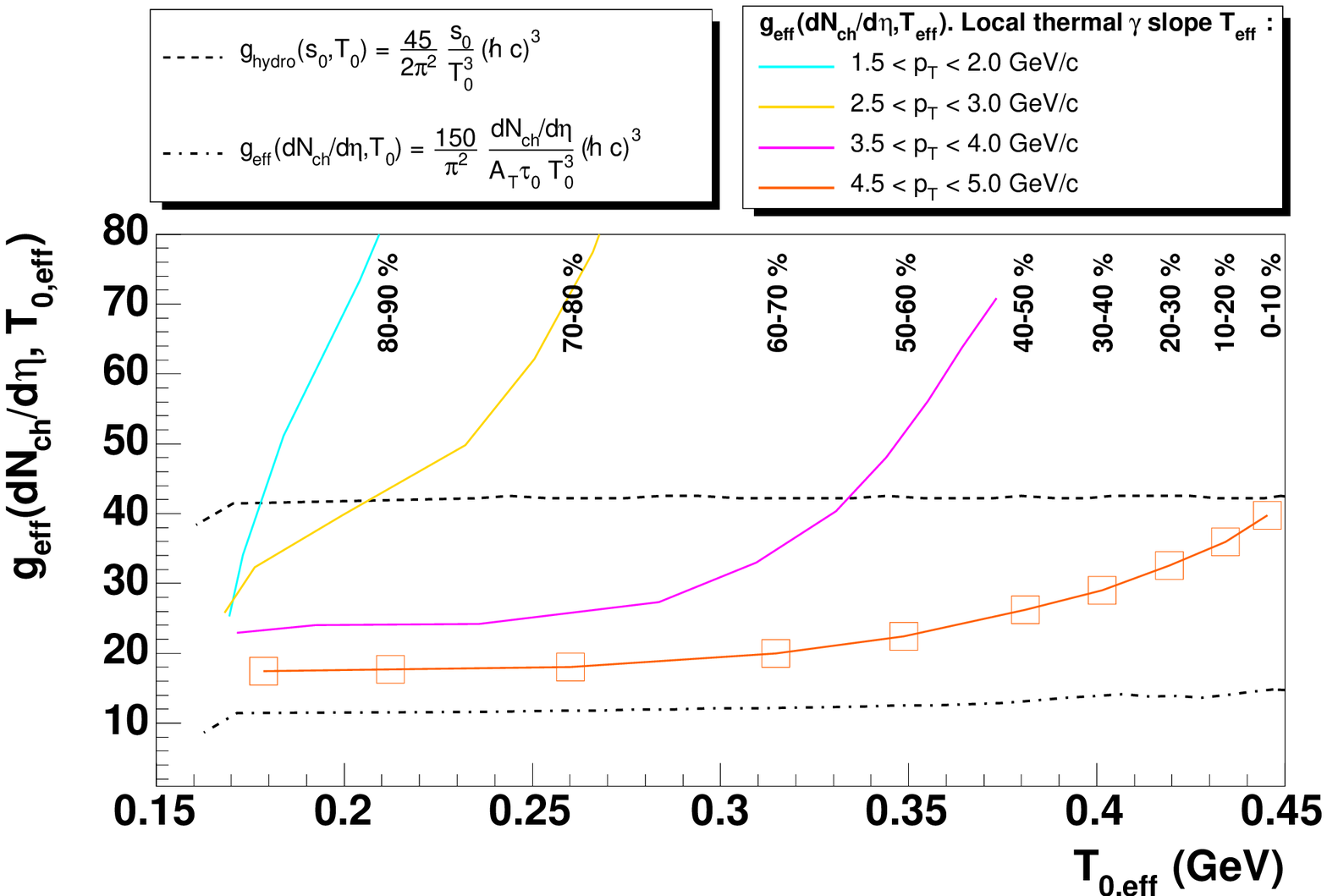,height=6.cm,width=9.cm}
\includegraphics
[height=4.5cm,width=9.cm,clip=true,viewport=0 0 567 310]{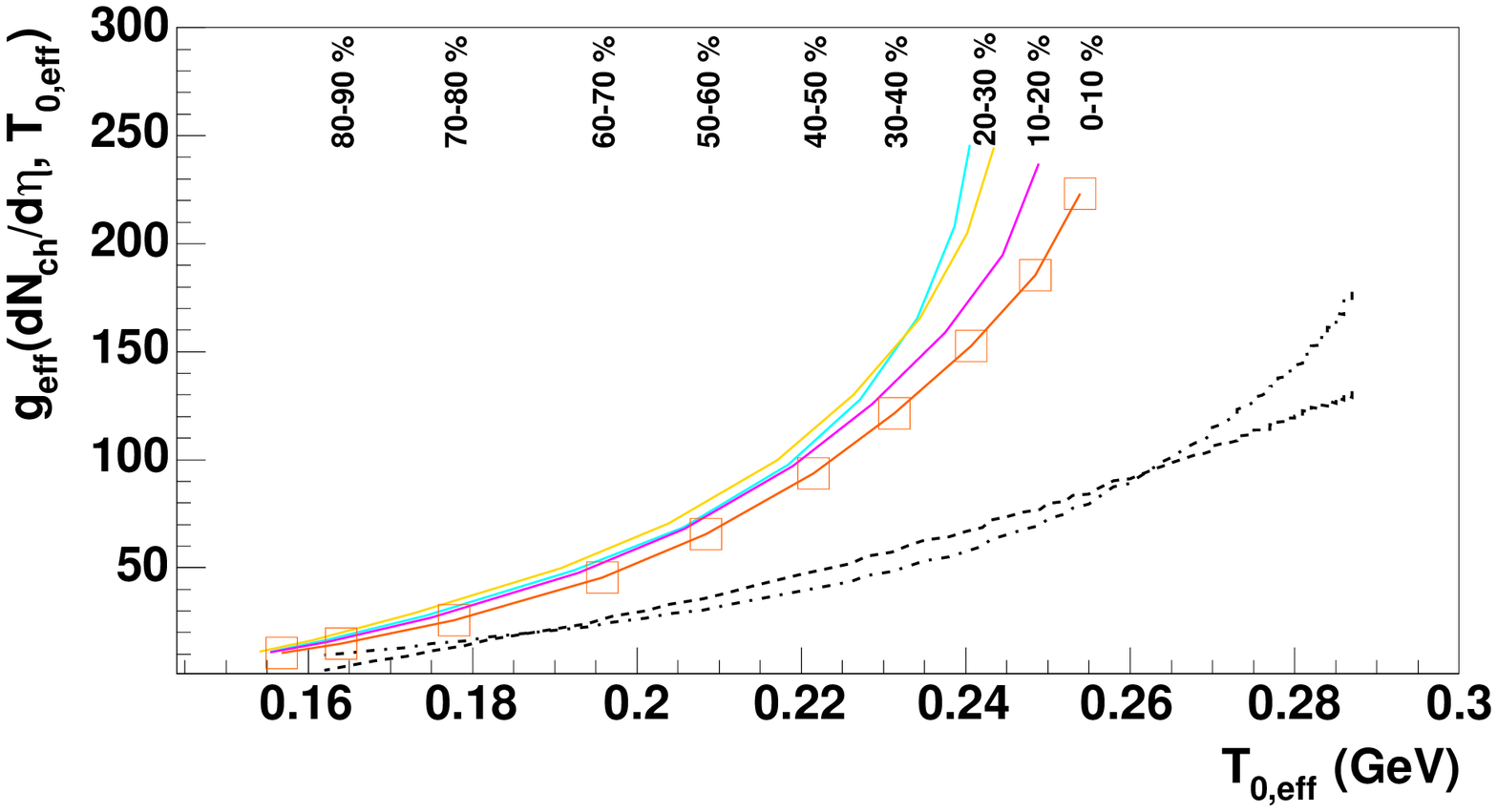}
\end{center}
\caption{Effective initial number of degrees of freedom obtained from our
hydrodynamical calculations with a QGP+HRG EoS (upper plot), and with a
pure HRG EoS (bottom), plotted as a function of the temperature ($T_{0}$) or 
thermal photon slope ($T_{\ensuremath{\it eff}}$) in different Au+Au centrality 
classes at $\sqrtsnn$ = 200 GeV. The number of degrees of freedom are
computed respectively: (i) From our initial thermodynamical conditions
$(s_0,\,T_0)$ via Eq.~(\protect\ref{eq:ndf}) (dashed line), 
(ii) from the obtained charged hadron multiplicity 
$dN_{ch}/d\eta$ and the {\it true} initial temperature $T_0$
via Eq.~(\protect\ref{eq:geff}) (dotted-dashed line); and
(iii) from $dN_{ch}/d\eta$ and the thermal photon slopes
$T_{\ensuremath{\it eff}}$ measured in different $p_T$ ranges
via Eq.~(\protect\ref{eq:geff}) (solid lines). For illustrative purposes,
the open squares indicate the approximate position of the different Au+Au 
centrality classes (in 10\% percentiles) for the values of $g_{\ensuremath{\it eff}}$ 
obtained using the thermal photon slopes measured above $p_T$ = 4 GeV/$c$.}
\label{fig:EoS}
\end{figure}

We propose here to use $T_{\ensuremath{\it eff}}$ as a proxy for the initial 
temperature of the system, and directly study the evolution, versus $T_{\ensuremath{\it eff}}$,
of the effective number of degrees of freedom defined as\footnote{Units are in GeV and fm. 
$\zeta(4)$ = $\pi^4/90$, where $\zeta(n)$ is the Riemann zeta function.}
\begin{equation}
g(s,\,T)\,=\,\frac{\pi^2}{4\,\zeta(4)}\frac{s}{T^3}\,(\hbar
c)^3=\,\frac{45}{2 \pi^2}\,\frac{s}{T^3}(\hbar c)^3,
\label{eq:ndf}
\end{equation}
which coincides with the degeneracy of a weakly interacting gas of massless particles.
[In a similar avenue, B.~Muller and K.~Rajagopal~\cite{muller_rajagopal05} 
have recently proposed a method to estimate the number of thermodynamic 
degrees of freedom via $g_{\ensuremath{\it eff}}\propto s^4/\epsilon^3$, where $s$ is 
also determined from the final hadron multiplicities].
The dashed line in Fig.~\ref{fig:EoS} (top) shows the evolution of the {\it true}
number of degrees of freedom $g_{\ensuremath{\it hydro}}(s_0,T_0)$ computed via
Eq.~(\ref{eq:ndf}), as a function of the (maximal) temperatures and entropies
directly obtained from the initial conditions of our hydrodynamical model in different 
Au+Au centralities\footnote{In the most peripheral reactions, the bag entropy has been 
subtracted to make more apparent the drop near $T_c$.}.
The first thing worth to note is that $g(s,T)$ remains constant at the expected 
degeneracy $g_{\ensuremath{\it hydro}}$ = 42.25 of an ideal gas of
$N_f$ = 2.5 quarks and gluons for basically {\it all} the
maximum temperatures accessible in the different centralities of Au+Au at
$\sqrt{s_{\mbox{\tiny{\it{NN}}}}}$ = 200 GeV. This indicates that
at top RHIC energies and for most of the impact parameters, $T_0$ is (well)
above $T_{crit}$ and the hottest parts of the initial fireball are in the QGP phase.
The expected drop in $g_{\ensuremath{\it hydro}}$ related to the transition
to the hadronic phase is only seen, if at all, for the very most peripheral
reactions (with $T_0\approx T_c$). 
Thus, direct evidence of the QGP-HRG phase change itself via the study of the centrality 
dependence of any experimentally accessible observable would only be potentially feasible 
at RHIC in Au+Au reactions at {\it lower} center-of-mass energies~\cite{dde_dima}.\\

As aforementioned, we can empirically trace the QCD EoS shown in Fig.~\ref{fig:EoS} 
(and eventually determine the temperature-evolution of the thermodynamic degrees of freedom 
of the produced medium) using the estimate of the initial temperature given by the 
thermal photon slopes, $T_{\ensuremath{\it eff}}$, and a second observable closely 
related to the initial entropy of the system such as the final-state hadron multiplicity, $dN/dy$.
Although one could have also considered to obtain $g_{\ensuremath{\it eff}}$ via 
$\varepsilon/T^4 \propto (dE_{T}/dy)/T_{\ensuremath{\it eff}}^4$,
using the transverse energy per unit rapidity $dE_{T}/dy$ measured in different Au+Au 
centralities~\cite{ppg019}, we prefer to use the expression (\ref{eq:ndf}) which contains
the entropy-, rather than the energy-, density for two reasons:
\begin{description}
\item (i) the experimentally accessible values of $dN/dy$ remain constant in an isentropic expansion 
(i.e. $dN/dy \propto s_0$) whereas, due to longitudinal work, the measured final
$dE_{T}/dy$ provides only a {\it lower limit} on the initial $\varepsilon$ 
($dE_{T}/dy \lesssim \varepsilon_0$); and
\item (ii) $g_{\ensuremath{\it eff}}\propto s/T_{\ensuremath{\it eff}}^3$ is less sensitive to 
experimental uncertainties associated to the measurement of $T_{\ensuremath{\it eff}}$ than 
$g_{\ensuremath{\it eff}}\propto \varepsilon/T_{\ensuremath{\it eff}}^4$ is.
\end{description}

\begin{sloppypar}
Again, in the absence of dissipative effects, the space-time evolution of the 
produced system in a nucleus-nucleus reaction is isentropic and the entropy density 
(per unit rapidity) at the thermalization time $\tau_{0}$ can be directly connected 
(via $s \,\approx\, 4 \,\rho$~\cite{wong_book}) to the final charged hadron 
pseudo-rapidity density\footnote{This formula uses $N_{tot}/N_{ch}$ = 3/2, 
and the Jacobian $|d\eta/dy|=E/p\approx$ 1.2.}:
\begin{equation}
s\,\approx\,4\cdot\frac{dN}{dV}\,\approx\,
\frac{7.2}{\mean{A_\perp}\cdot\tau_0}\cdot\frac{dN_{ch}}{d\eta}
\label{eq:entropy}
\end{equation}
where we have written the volume of the system, $dV=\mean{A_\perp}\tau_0\,d\eta$,
as the product of the (purely geometrical) average transverse overlap area for 
each centrality times the starting proper time of our hydro evolution ($\tau_0=0.15$ fm/$c$),
and where $dN_{ch}/d\eta$ is the {\it charged} hadron multiplicity customarily
measured experimentally at mid-rapidity\footnote{Note again that both the
photon slopes and the charged hadron multiplicities are proxies of the thermodynamical
conditions of the system {\it at the same time} $\tau_{0}$.}.
By combining, Eqs.~(\ref{eq:ndf}) and  (\ref{eq:entropy}), 
we obtain an estimate for the number of degrees of freedom of the system 
produced in a given A+A collision at impact parameter $b$:
\begin{equation}
g_{\ensuremath{\it eff}}\left(\frac{dN_{ch}(b)}{d\eta},\,T_{\ensuremath{\it eff}}(b)\right)\,\approx
\,\frac{150}{\pi^2}\cdot\frac{(\hbar c)^3}{\mean{A_\perp(b)}\cdot\tau_0\cdot T^{3}_{\ensuremath{\it eff}}(b)}\cdot
\frac{dN_{ch}(b)}{d\eta}\;,
\label{eq:geff}
\end{equation}
which can be entirely determined with two experimental observables: $dN_{ch}/d\eta$ and 
$T_{\ensuremath{\it eff}}$.\\
\end{sloppypar}

Let us first assess to what extent the ansatz~(\ref{eq:geff}) is affected by the assumption that
Eq.~(\ref{eq:entropy}) indeed provides a good experimental measure of the initial entropy 
density $s$. The dotted-dashed curve in Fig.~\ref{fig:EoS} has been obtained via
Eq.~(\ref{eq:geff}) using the $(dN_{ch}/d\eta)/\mean{A_\perp}$ values obtained from 
our hydrodynamical model, and the {\it true} (input) initial temperature of the system $T_0$, 
and thus it is only sensitive to the way we estimate the entropy density. The resulting curve 
is a factor of $\sim$3 below the expected ``true'' $g_{\ensuremath{\it hydro}}$ curve, i.e.
$g_{\ensuremath{\it eff}}\left(dN_{ch}/d\eta,T_{0}\right)\approx 3\cdot g_{\ensuremath{\it hydro}}(s_0,T_0)$,
indicating that Eq.~(\protect\ref{eq:geff}) underestimates by the same amount the maximal entropy of the 
original medium. This is so because our estimate $(dN_{ch}/d\eta)/\mean{A_\perp}$ specifies the entropy 
density averaged over the {\it whole} Glauber transverse area $\mean{A_\perp}$, whereas the maximal 
entropy area in the {\it core} of the system (from where the hardest thermal photons are emitted) 
is $\sim$3 times {\it smaller}. Although one could think of a method to correct for this difference, 
this would introduce an extra model-dependence that we want to avoid at this point. 
We prefer to maintain the simple (geometrical overlap) expression of the transverse area
$\mean{A_\perp(b)}$ in Eq.~(\protect\ref{eq:geff}), and exploit the fact that, although such an 
equation does not provide the true {\it absolute} number of degrees of freedom, it does provide a 
very reliable indication of the dependence of $g_{\ensuremath{\it eff}}$ on the temperature
of the system and, therefore, of the exact {\it form} of the underlying EoS.\\

Finally, let us consider the last case where we use Eq.~(\ref{eq:geff}) with 
the values of $dN_{ch}/d\eta$ {\it and} $T_{\ensuremath{\it eff}}$ that can be actually 
experimentally measured. The different solid curves in the upper plot of Fig.~\ref{fig:EoS} 
show the effective degeneracy $g_{\ensuremath{\it eff}}$, computed using
Eq.~(\ref{eq:geff}) and the local photon slopes $T_{\ensuremath{\it eff}}$
measured in different $p_T$ ranges for our default QGP+HRG evolution.
As one could expect from Fig.~\ref{fig:thermal_slopes}, the best reproduction
of the shape of the underlying EoS 
is obtained with the effective temperatures measured in higher $p_T$ bins. 
For those $T_{\ensuremath{\it eff}}$, the computed $g_{\ensuremath{\it eff}}$'s 
show a relatively constant value in a wide range of centralities as expected for
a weakly interacting QGP. Deviations from this ideal-gas plateau appear
for more central collisions, due to an increasing difference between the
(high) initial temperatures, $T_0$, and the apparent
temperature given by the photon slopes (Fig.~\ref{fig:thermal_slopes}).
Such deviations do not spoil, however, the usefulness of our estimate
since, a non-QGP EoS would result in a considerably different
dependence of $g_{\ensuremath{\it eff}}$ on the reaction centrality. Indeed,
the different curves in the bottom plot of Fig.~\ref{fig:EoS} obtained with
a pure hadron resonance gas EoS clearly indicate\footnote{Accidentally,
$g_{\ensuremath{\it eff}}\gtrsim g_{\ensuremath{\it hydro}}$ in the case of a HRG EoS,
because the underestimation of the apparent temperature (raised to the cube)
compensates for the aforementioned area averaging of the entropy.}
that a HRG EoS, or in general any EoS with exponentially increasing number of mass states, 
would bring about a much more dramatic rise of $g_{\ensuremath{\it eff}}$ with $T_{\ensuremath{\it eff}}$.\\

In summary, the estimate~(\ref{eq:geff}) indeed provides a direct experimental handle 
on the {\it form} of the EoS of the strongly interacting medium produced in the first 
instants of high-energy nuclear collisions. More quantitative conclusions on the possibility
to extract the exact shape of the underlying EoS and/or the absolute number of degrees of freedom 
of the produced medium require more detailed theoretical studies (e.g. with varying 
lattice-inspired EoS's~\cite{dde_dima} and/or using more numerically involved 3D+1 hydrodynamical approaches).
In any case, we are confident that by experimentally measuring the thermal photon slopes
in different Au+Au centralities and correlating them with the associated charged 
hadron multiplicities as in Eq.~(\ref{eq:geff}), one can approximately observe the 
expected ``plateau'' in the number of degrees of freedom indicative of QGP formation 
above a critical value of $T$.

\section{Conclusions}
We have studied thermal photon production in Au+Au reactions at
$\sqrt{s_{\mbox{\tiny{\it{NN}}}}}$ = 200 GeV using a Bjorken hydrodynamic
model with longitudinal boost invariance. We choose the initial conditions
of the hydrodynamical evolution so as to efficiently reproduce the
observed particle multiplicity in central Au+Au collisions at RHIC
and use a simple Glauber prescription to obtain the corresponding initial
conditions for all other centralities. With such a model we can perfectly 
reproduce the identified soft pion, kaon and proton $p_T$-differential spectra measured at RHIC. 
Complementing our model with the most up-to-date parametrizations of the QGP and 
HRG thermal photon emission rates plus a NLO pQCD calculation of the
prompt $\gamma$ contribution, we obtain direct photon spectra which 
are in very good agreement with the Au+Au direct photon (upper limit) yields
measured by the PHENIX experiment. In central collisions, a thermal
photon signal should be identifiable as a factor of $\sim$8 -- 1 excess
over the pQCD $\gamma$ component within $p_T\approx$ 1 -- 4 GeV/$c$,
whereas pure prompt emission clearly dominates the photon spectra
at all $p_T$ in peripheral reactions. The local inverse slope parameter of the 
thermal photon spectrum is found to be directly correlated to the maximum temperature 
attained in the course of the collision. The experimental measurement of local thermal photon 
slopes above $p_T\approx$ 2.5 GeV/$c$, with values $T_{\ensuremath{\it eff}}\gtrsim$ 250 MeV
and with pronounced $p_T$ dependences can only be reproduced
by space-time evolutions of the reaction that include a QGP phase.\\

Finally, we have proposed and tested within our framework, an empirical method to 
determine the effective thermodynamical number of degrees of freedom of the produced medium,
$g(s,T)\propto s(T)/T^3$, by correlating the thermal photon slopes with the final-state charged 
hadron multiplicity measured in different centrality classes. We found that one can 
clearly distinguish between the equation of state of a weakly interacting
quark-gluon plasma and that of a system with rapidly rising number of mass
states with $T$. Stronger quantitative conclusions on the exact shape of the 
underlying EoS and/or the absolute number of degrees of freedom 
of the produced medium require more detailed theoretical studies as well as high
precision photon data in Au+Au and baseline p+p, d+Au collisions. 
In any case, the requirement for hydrodynamical models of 
concurrently describing the experimental bulk hadron and thermal photon spectra for 
different Au+Au centralities at $\sqrt{s_{\mbox{\tiny{\it{NN}}}}}$ = 200 GeV, imposes
very strict constraints on the form of the equation of state of the underlying
expanding QCD matter produced in these reactions.

\section{Acknowledgments}

We would like to thank Werner Vogelsang for providing us with his NLO
pQCD calculations for photon production in p+p collisions at
$\sqrt{s}$ = 200 GeV; Sami Rasanen for valuable comments on hydrodynamical
photon production; and Helen Caines and Olga Barannikova for useful discussions
on (preliminary) STAR hadron data. D.P. acknowledges support from
MPN of Russian Federation under grant NS-1885.2003.2.




\begin{thebibliography}{00}

\def\IJMPA{{Int. J. Mod. Phys.}~{\bf A}}
\def\EPJ{{Eur. Phys. J.}~{\bf C}}
\def\JPG{{J. Phys}~{\bf G}}
\def\JHEP{{J. High Energy Phys.}~}
\def\NCA{Nuovo Cimento~}
\def\NIM{Nucl. Instrum. Methods~}
\def\NIMA{{Nucl. Instrum. Methods}~{\bf A}}
\def\NPA{{Nucl. Phys.}~{\bf A}}
\def\NPB{{Nucl. Phys.}~{\bf B}}
\def\PLB{{Phys. Lett.}~{\bf B}}
\def\PLC{Phys. Repts.\ }
\def\PRL{Phys. Rev. Lett.\ }
\def\PRD{{Phys. Rev.}~{\bf D}}
\def\PRC{{Phys. Rev.}~{\bf C}}
\def\ZPC{{Z. Phys.}~{\bf C}}

\bibitem{latt}See e.g. F.~Karsch, {\it Lect. Notes Phys.} {\bf 583}, 209 (2002).
\bibitem{sps_qgp}U.W.~Heinz and M.~Jacob, nucl-th/0002042.
\bibitem{rhic_qgp}M.~Gyulassy and L.~McLerran, \NPA {\bf 750}, 30 (2005).
\bibitem{feinberg}E.~L.~Feinberg, Nuovo Cim.\ A {\bf 34}, 391 (1976).
\bibitem{shuryak_photons}E.~V.~Shuryak, \PLB {\bf 78}, 150 (1978)
[Sov.\ J.\ Nucl.\ Phys.\  {\bf 28} (1978\ YAFIA,28,796-808.1978) 408.1978\ YAFIA,28,796].
\bibitem{peitz_thoma_physrep}T.~Peitzmann and M.~H.~Thoma, Phys.\ Rept.\  {\bf 364}, 175 (2002).
\bibitem{yellow_rep}F.~Arleo {\it et al.}, in ``CERN Yellow Report on Hard Probes in Heavy Ion Collisions at the LHC'';
hep-ph/0311131.
\bibitem{gale_rep}C.~Gale and K.~L.~Haglin, in ``Quark Gluon Plasma. Vol 3''
Eds: R.C. Hwa and X.N. Wang, World Scientific, Singapore, hep-ph/0306098.
\bibitem{wa98_photons}M.~M.~Aggarwal {\it et al.} [WA98 Collaboration], \PRL {\bf 85}, 3595 (2000).
\bibitem{srivastava_sps_rhic}D.~K.~Srivastava and B.~Sinha, \PRC {\bf 64}, 034902 (2001); 
D.~K.~Srivastava, Pramana {\bf 57}, 235 (2001).
\bibitem{alam_sps_rhic}J.~e.~Alam, S.~Sarkar, T.~Hatsuda, T.~K.~Nayak and B.~Sinha, \PRC {\bf 63}, 021901 (2001).
\bibitem{peressou}D.~Y.~Peressounko and Y.~E.~Pokrovsky, \NPA {\bf 624}, 738 (1997); 
\NPA {\bf 669}, 196 (2000); 
hep-ph/0009025.
\bibitem{steffen_sps_rhic_lhc}F.~D.~Steffen and M.~H.~Thoma, \PLB {\bf 510}, 98 (2001).
\bibitem{finnish_hydro}P.~Huovinen, P.~V.~Ruuskanen and S.~S.~Rasanen, \PLB {\bf 535}, 109 (2002);
S.~S.~Rasanen, \NPA {\bf 715}, 717 (2003); 
H.~Niemi, S.~S.~Rasanen and P.V.~Ruuskanen in~\cite{yellow_rep}.
\bibitem{dde_sps}D.~d'Enterria, \PLB 596, 32 (2004).
\bibitem{aurenche}P.~Aurenche, M.~Fontannaz, J.~P.~Guillet, B.~A.~Kniehl, E.~Pilon and M.~Werlen, \EPJ {\bf 9} (1999) 107.
\bibitem{wong}C.~Y.~Wong and H.~Wang, \PRC {\bf 58}, 376 (1998).
\bibitem{apanasevich} L.~Apanasevich {\it et al.}, \PRD {\bf 63}, 014009 (2001).
\bibitem{cronin}J.W.~Cronin {\it et al.}, \PRD {\bf 11}, 3105 (1975); 
D.~Antreasyan {\it et al.}, \PRD {\bf 19}, 764 (1979).
\bibitem{dumitru}A.~Dumitru and N.~Hammon, hep-ph/9807260; 
A.~Dumitru, L.~Frankfurt, L.~Gerland, H.~Stocker and M.~Strikman, \PRC {\bf 64}, 054909 (2001).
\bibitem{ina}S.~Jeon, J.~Jalilian-Marian and I.~Sarcevic, \NPA {\bf 715}, 795 (2003).
\bibitem{levai}G.~Papp, G.~I.~Fai and P.~Levai, hep-ph/9904503.
\bibitem{ppg042}J.~Frantz [PHENIX Collaboration], \JPG {\bf 30}, S1003 (2004); 
S.S.~Adler {\it et al.} [PHENIX Collaboration], \PRL {\bf 94} 232301 (2005).
\bibitem{ppg049}K.~Okada [PHENIX Collaboration], {\it Proceeds. SPIN'04}, hep-ex/0501066; 
S.S.~Adler {\it et al.} [PHENIX Collaboration], \PRD {\bf 71} 071102 (2005).
\bibitem{ppg024}S.S.~Adler {\it et al.} [PHENIX Collaboration], \PRL {\bf 91}, 241803 (2003).
\bibitem{dAu_phnx}S.~S.~Adler {\it et al.} [PHENIX Collaboration], \PRL  {\bf 91}, 072303 (2003);
H.~Buesching [PHENIX Collaboration], nucl-ex/0410002.
\bibitem{kolb_heinz_rep}P.~F.~Kolb and U.~Heinz, in ``Quark Gluon Plasma. Vol 3'' 
Eds: R.C. Hwa and X.N. Wang, World Scientific, Singapore, nucl-th/0305084, and refs. therein.
\bibitem{teaney_hydro}D.~Teaney, J.~Lauret and E.~V.~Shuryak, nucl-th/0110037.
\bibitem{hirano}T.~Hirano and K.~Tsuda, \PRC {\bf 66}, 054905 (2002); 
T.~Hirano and Y.~Nara, \PRC {\bf 69}, 034908 (2004).
\bibitem{bjorken}J.~D.~Bjorken, \PRD {\bf 27}, 140 (1983).
\bibitem{brahms_hadrons}I.~G.~Bearden {\it et al.} [BRAHMS Collaboration], \PRL {\bf 94} 162301 (2005); 
I.~Arsene {\it et al.} [BRAHMS Collaboration],\PRC {\bf 72} 014908 (2005).
\bibitem{eskola_hydro}K.~J.~Eskola, K.~Kajantie and P.~V.~Ruuskanen, \EPJ {\bf 1}, 627 (1998).
\bibitem{mohanty_cs}B.~Mohanty and J.~e.~Alam, \PRC {\bf 68}, 064903 (2003).
\bibitem{maccormack}R.~W.~MacCormack and A.~J.~Paullay, Computers and Fluids, Vol. 2, 1974 (Oxford: Pergamon)
\bibitem{pbm_thermal}P.~Braun-Munzinger, K.~Redlich and J.~Stachel, in ``Quark Gluon Plasma. Vol 3'' 
Eds: R.C. Hwa and X.N. Wang, World Scientific, Singapore, nucl-th/0304013.
\bibitem{andronic05}A.~Andronic, P.~Braun-Munzinger and J.~Stachel, nucl-th/0511071.
\bibitem{cooper_frye}F.~Cooper and G.~Frye, \PRD {\bf 10}, 186 (1974).
\bibitem{PDG}K.~Hagiwara {\it et al.} [Particle Data Group Collaboration], 
\PRD {\bf 66}, 010001 (2002).
\bibitem{hahn}B.~Hahn, D.G.~Ravenhall and R.~Hofstadter, {Phys. Rev.}~{\bf 101}, 1131 (1956).
\bibitem{kolb_heinz_finnishgroup}P.~F.~Kolb, U.~W.~Heinz, P.~Huovinen, K.~J.~Eskola and K.~Tuominen, \NPA {\bf 696}, 197 (2001).
\bibitem{ppg019} S.~S.~Adler {\it et al.} [PHENIX Collaboration], \PRC {\bf 71}, 034908 (2005)  
[Erratum-ibid.\ C {\bf 71}, 049901 (2005)].
\bibitem{star_Nch}T.~S.~Ullrich [STAR Collaboration], Heavy Ion Phys.\  {\bf 21}, 143 (2004).
\bibitem{phobos_Nch}B.~B.~Back {\it et al.} [PHOBOS collaboration], \NPA {\bf 715}, 65 (2003).
\bibitem{brahms_Nch}I.~G.~Bearden {\it et al.} [BRAHMS Collaboration], \PRL {\bf 88}, 202301 (2002).
\bibitem{vogel_hadrons}F.~Aversa {\it et al.}, \NPB {\bf 327}, 105 (1989); 
B.~Jager {\it et al.} \PRD {\bf 67}, 054005 (2003); 
W.~Vogelsang, private communication (NLO spectra have been computed
with CTEQ6 PDFs, KKP FFs, and scales set to the hadron $p_T$).
\bibitem{frankfurt_rhic_lhc}N.~Hammon, A.~Dumitru, H.~Stoecker and W.~Greiner, \PRC {\bf 57}, 3292 (1998).
\bibitem{bass}S.~A.~Bass, B.~Muller and D.~K.~Srivastava, \PRL {\bf 90}, 082301 (2003).
\bibitem{bass2}T.~Renk, S.~A.~Bass and D.~K.~Srivastava, nucl-th/0505059.
\bibitem{berges04}J.~Berges, S.~Borsanyi and C.~Wetterich, \PRL  {\bf 93}, 142002 (2004).
\bibitem{arnold04}P.~Arnold, J.~Lenaghan, G.~D.~Moore and L.~G.~Yaffe, nucl-th/0409068.
\bibitem{kolb_rapp_flow} P.~F.~Kolb and R.~Rapp, \PRC {\bf 67}, 044903 (2003).
\bibitem{ppg026}S.~S.~Adler {\it et al.} [PHENIX Collaboration], \PRC {\bf 69}, 034909 (2004).
\bibitem{star_hadrons}J.~Adams {\it et al.} [STAR Collaboration], \PRL  {\bf 92}, 112301 (2004).
\bibitem{star_hadrons2}H.~Caines [STAR Collaboration], \JPG {\bf 31} (2005) S101; 
O.~Barannikova [STAR Collaboration], nucl-ex/0408022 and private communication.
\bibitem{phobos_lowpt_had}B.B. Back {\it et al.} [PHOBOS Collaboration], \PRC {\bf 70}, 051901(R) (2004).
\bibitem{phenix_hiptpi0_200}S.~S.~Adler {\it et al.} [PHENIX Collaboration], \PRL {\bf 91}, 072301 (2003).
\bibitem{star_hipt_200}J.~Adams {\it et al.} [STAR Collaboration], \PRL {\bf 91}, 172302 (2003).
\bibitem{phnx_ppbar}K.~Adcox {\it et al.} [PHENIX Collaboration], \PRL  {\bf 88}, 242301 (2002); 
S.~S.~Adler {\it et al.} [PHENIX Collaboration], \PRL  {\bf 91}, 172301 (2003).
\bibitem{recomb}R.~J.~Fries, \JPG {\bf 30}, S853 (2004) and refs. therein.
\bibitem{vogel_gamma}L.E.~Gordon and W.~Vogelsang, \PRD {\bf 48}, 3136 (1993); 
\PRD {\bf 50}, 1901 (1994); 
and W.~Vogelsang (private communication); also P.~Aurenche {\it et al.}, \PLB {\bf 140}, 87 (1984); 
and \NPB {\bf 297}, 661 (1988).
\bibitem{cteq6}S.~Kretzer, H.~L.~Lai, F.~I.~Olness and W.~K.~Tung, \PRD {\bf 69}, 114005 (2004).
\bibitem{grv_photons}M.~Gluck, E.~Reya and A.~Vogt, \PRD {\bf 48}, 116 (1993)
[Erratum-ibid.\ D {\bf 51} (1995) 1427].
\bibitem{jamal}J.~Jalilian-Marian, K.~Orginos and I.~Sarcevic, \PRC {\bf 63}, 041901 (2001).
\bibitem{qm05}S.~Bathe [PHENIX Collaboration], nucl-ex/0511042, Quark Matter'05, Budapest, August 2005.
\bibitem{arleo04}F.~Arleo, hep-ph/0406291.
\bibitem{turbide05}S.~Turbide, C.~Gale, S.~Jeon and G.~D.~Moore, \PRC {\bf 72}, 014906 (2005).
\bibitem{dde_hq04}D.~d'Enterria, \JPG {\bf 31}, S491 (2005).
\bibitem{zakharov04}B.~G.~Zakharov, JETP Lett.\  {\bf 80}, 1 (2004) [Pisma Zh.\ Eksp.\ Teor.\ Fiz.\  {\bf 80}, 3 (2004)].
\bibitem{arnold}P.~Arnold, G.~D.~Moore and L.~G.~Yaffe, JHEP {\bf 0112}, 009 (2001).
\bibitem{traxler}C.~T.~Traxler, H.~Vija and M.~H.~Thoma, \PLB {\bf 346}, 329 (1995).
\bibitem{gelis}F.~Gelis, H.~Niemi, P.~V.~Ruuskanen and S.~S.~Rasanen, \JPG {\bf 30}, S1031 (2004).
\bibitem{turbide}S.~Turbide, R.~Rapp and C.~Gale, \PRC {\bf 69}, 014903 (2004).
\bibitem{karsch_alphaS}O.~Kaczmarek, F.~Karsch, F.~Zantow and P.~Petreczky, \PRD {\bf 70}, 074505 (2004).
\bibitem{fries_rhic_lhc}R.~J.~Fries, B.~Muller and D.~K.~Srivastava, \PRL {\bf 90}, 132301 (2003).
\bibitem{fries_gammajet2}R.~J.~Fries, B.~Muller and D.~K.~Srivastava, nucl-th/0507018.
\bibitem{arleo05}F.~Arleo, hep-ph/0601075.
\bibitem{alam05}J.~e.~Alam, J.~K.~Nayak, P.~Roy, A.~K.~Dutt-Mazumder and B.~Sinha, nucl-th/0508043.
\bibitem{qm05_akiba}Y.~Akiba [PHENIX Collaboration], nucl-ex/0510008, Quark Matter'05, Budapest, August 2005.
\bibitem{muller_rajagopal05}B.~Muller and K.~Rajagopal,\EPJ {\bf 43} (2005) 15.
\bibitem{dde_dima}D.~d'Enterria and D.~Peressounko, in preparation.
\bibitem{wong_book}See e.g. C.~Y.~Wong, \textit{Introduction to high-energy heavy ion collisions}, Singapore, 
World Scientific (1994). Note that the relation $s \,=\, 4\,\zeta(4)/\zeta(3)\,\rho\,\approx\, 3.6 \,\rho$
commonly found in the literature refers to an ideal gas of bosons. For an ideal gas of quarks and gluons
(with $N_{B,F}$ bosons and fermions), the effective number of degrees of freedom 
for $\rho$ ($g_{\ensuremath{\it eff}} = N_B + 3/4\cdot N_F$) and for $\epsilon$
($g_{\ensuremath{\it eff}} = N_B + 7/8\cdot N_F$, which ``propagate'' to the 
entropy density via: $s = (\epsilon+P)/3$) are slightly different. 
Thus, the right expression is $s \,\approx\, 4 \,\rho$.
\end{thebibliography}
\end{document}